\documentclass[twocolumn,epjc3]{svjour3}          

\RequirePackage[T1]{fontenc}

\smartqed  

\RequirePackage{amsmath}
\RequirePackage{graphicx}
\RequirePackage{mathptmx}      
\RequirePackage{flushend}
\usepackage{amsfonts}
\usepackage{ulem}
\RequirePackage{placeins}
\RequirePackage{mathtools}
\RequirePackage{morefloats}
\RequirePackage{tablefootnote}
\RequirePackage[labelfont=normalsize,textfont=normalsize]{subcaption}
\RequirePackage{tikz}
\usepackage{soul} 
\RequirePackage{slashed}
\usetikzlibrary{shapes,arrows,chains,calc}
\RequirePackage[numbers,sort&compress]{natbib}
\RequirePackage[breaklinks=true,colorlinks,citecolor=blue,urlcolor=blue,linkcolor=blue]{hyperref}
\journalname{Eur. Phys. J. C}

\begin{document}

\title{SCYNet: Testing supersymmetric models at the LHC with neural networks}


\author{Philip Bechtle\thanksref{e1,addr1}
        \and
        Sebastian Belkner\thanksref{e2,addr1}
        \and 
        Daniel Dercks\thanksref{e3,addr2}
        \and
        Matthias Hamer\thanksref{e4,addr1}
        \and
        Tim Keller\thanksref{e5,addr3}
        \and
        Michael Kr{\"a}mer\thanksref{e6,addr3}
        \and 
        Bj{\"o}rn Sarrazin\thanksref{e7,addr3}
        \and
        Jan Sch{\"u}tte-Engel\thanksref{e8,addr3}
        \and
        Jamie Tattersall\thanksref{e9,addr3}
}

\thankstext{e1}{bechtle@physik.uni-bonn.de}
\thankstext{e2}{sbelkner@th.physik.uni-bonn.de}
\thankstext{e3}{daniel.dercks@desy.de}
\thankstext{e4}{Matthias.Hamer@cern.ch}
\thankstext{e5}{tim.keller@rwth-aachen.de}
\thankstext{e6}{mkraemer@physik.rwth-aachen.de}
\thankstext{e7}{sarrazin@physik.rwth-aachen.de}
\thankstext{e8}{schuette@physik.rwth-aachen.de}
\thankstext{e9}{tattersall@physik.rwth-aachen.de}

\institute{Universit{\"a}t Bonn, Nussallee 12, Bonn, Germany\label{addr1}
          \and
          Universit{\"a}t Hamburg, Luruper Chaussee 149, Hamburg, Germany\label{addr2}
          \and
          Institute for Theoretical Particle Physics and Cosmology, RWTH Aachen University, 52074 Aachen, Germany\label{addr3}
}

\date{TTK-17-06}

\maketitle

\begin{abstract}

SCYNet (SUSY Calculating Yield Net) is a tool for testing
supersymmetric models against LHC data. It uses neural network 
regression for a fast evaluation of the profile likelihood ratio. 
Two neural network approaches 
have been developed: 
one network has been trained using the parameters of the 11-dimensional phenomenological 
Minimal Supersymmetric Standard Model (pMSSM-11) as an input and evaluates the
corresponding profile likelihood ratio within milliseconds. It can
thus be used in global 
pMSSM-11 fits without time penalty.
In the second approach, the neural network has been trained using
model-independent signature-related objects, such as energies and particle
multiplicities, which were estimated from the para\-me\-ters of a given new
physics model. While the calculation of the energies and particle
multiplicities takes up computation time, the corresponding neural network
is more general and can be used to predict the LHC
profile likelihood ratio for a wider class of new physics models. 

\end{abstract}

\section{Introduction}

Direct searches for new particles at the LHC are among the most
sensitive probes of beyond the Standard Model (BSM) physics and play a
crucial role in global BSM fits. 
Calculating the profile likelihood ratio (referred to as $\chi^2$ in
the following) for a new physics model from LHC searches is
straightforward in principle: for each point in the model parameter
space, signal events are generated using a Monte-Carlo simulation. The
$\chi^2$ is then calculated from the number of expected signal events,
the Standard Model background estimate and the number of observed
events for a given experimental signal region.  
The computation time for such simulations, can be overwhelming however, especially when testing BSM scenarios 
 with many model parameters. Global fits of supersymmetric (SUSY)
 models, for example, are typically based on the evaluation of ${\cal
 O}(10^9)$ parameter points, see e.g.\
 \cite{killingcmssm,deVries:2015hva,Strege:2014ija}, and the required Monte-Carlo
statistics for estimating the number of signal events for each
parameter point requires up to several hours of CPU time. In this
study we have attempted to provide a fast evaluation of the LHC $\chi^2$  for
generic SUSY models by utilizing neural network regression.

Global SUSY analyses which combine low-energy precision observables, like the anomalous
magnetic moment of the muon, and LHC searches for new particles
strongly disfavour minimal SUSY models, like the constrained Minimal Supersymmetric
Standard Model (cMSSM)~\cite{killingcmssm}. 
Thus, more general
supersymmetric models have to be explored, including for
example the phenomenological MSSM ($\mbox{pMSSM-11}$)~\cite{Djouadi:1998di},
specified by eleven SUSY parameters defined at the electroweak scale. The pMSSM-11 allows to accommodate the anomalous
magnetic moment of the muon, the dark matter relic density and the
LHC limits from direct searches. However, the large number of model
parameters poses a significant challenge for global pMSSM-11 fits.

\begin{figure*}[t]
\begin{subfigure}[c]{\textwidth}
\centering
\large
\subcaption{Direct approach}
\begin{tikzpicture}[%
    >=triangle 60,  
    start chain=going right,    
    node distance=4mm and 40mm, 
    every join/.style={norm},
    scale=0.8,
    every node/.style={transform shape},
    ]
\tikzset{
  base/.style={draw, on chain, on grid, align=center, minimum height=4ex},
  proc/.style={base, rectangle, align=center, text width=8em},
  proc2/.style={base, rectangle, align=center, text width=6em, minimum height=6em},
}
\node [proc] (p0) {pMSSM-11 Parameters};
\node [proc2, right=of p0, text centered] (p1) {\centering Neural network};
\node [proc, right=of p1] (p2) {$\chi^2_{\rm SN}$};

\draw[->] (p0.east) --  (p1.west);
\draw[->] (p1.east) -- (p2.west);

\end{tikzpicture}
\captionsetup{font=normalsize}
\\\vspace{5mm}
\end{subfigure}
\begin{subfigure}[c]{\textwidth}
\centering
\large
\subcaption{Reparametrized approach}
\begin{tikzpicture}[%
  >=triangle 60, start chain=going right, node distance=12mm and 40mm,
  every join/.style={norm}, scale=0.8, every node/.style={transform
    shape}, ]
\tikzset{
  base/.style={draw, on chain, on grid, align=center, minimum height=4ex},
  proc/.style={base, rectangle, align=center, text width=8em},
  proc2/.style={base, rectangle, align=center, text width=6em, minimum height=6em},
}
\node [proc] (p0) {pMSSM-11 parameters};
\node [proc, right=45mm of p0, text centered] (p11) {\centering Cross sections};
\node [proc, below=of p11, text centered] (p12) {\centering Branching ratios};
\node [proc, above=of p11, text centered] (p13) {\centering Masses};
\node [proc, right=45mm of p11, text centered] (p1) {\centering Reparametrization};
\node [proc2, right=of p1, text centered] (p2) {\centering Neural network};
\node [proc, right=of p2] (p3) {$\chi^2_{\rm SN}$};

\draw[->] (p0.east) --  (p11.west);
\draw[->] (p0.south east) --  (p12.west);
\draw[->] (p0.north east) --  (p13.west);

\draw[->] (p11.east) -- (p1.west);
\draw[->] (p12.east) -- (p1.south west);
\draw[->] (p13.east) -- (p1.north west);

\draw[->] (p1.east) -- (p2.west);

\draw[->] (p2.east) --  (p3.west);

\end{tikzpicture}
\end{subfigure}
\caption{\sl Flowchart indicating how we have calculated the LHC $\chi^2$ via the direct (a) and 
reparametrized (b) neural network regression approach described in the
text. Throughout the paper, the subscript ``SN'' denotes the $\chi^2$ obtained via
SCYNet.}
\label{fig:nn_flow} 
\end{figure*}
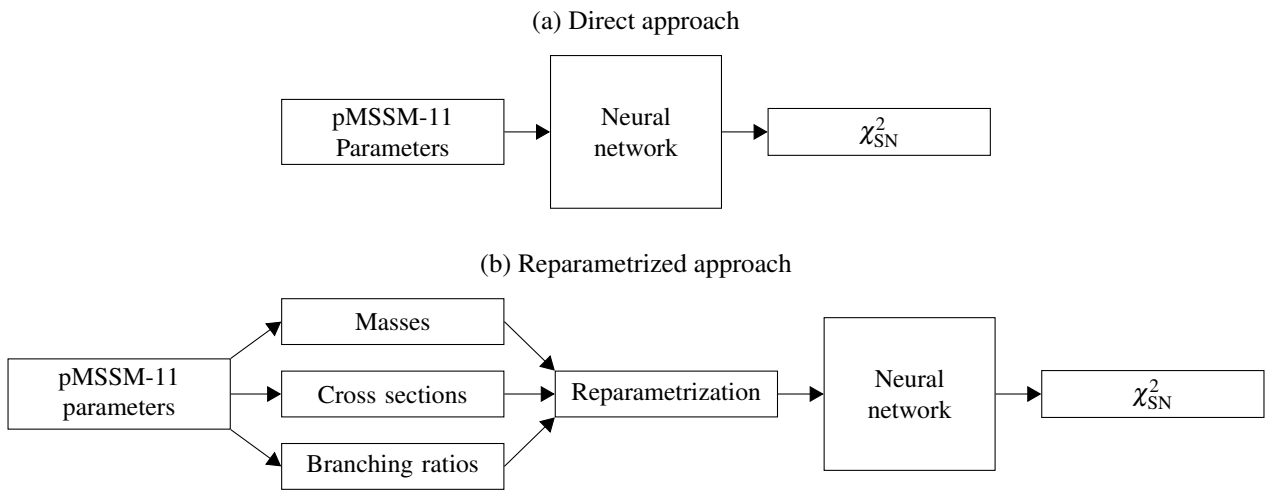

In this paper we introduce the SCYNet tool for the accurate and fast
statistical evaluation of 
LHC search limits -- and potential signals -- within the pMSSM-11 and
other SUSY models. SCYNet is based on the results of a simulation of
new physics models that used
CheckMATE~2.0~\cite{Drees:2013wra,Kim:2015wza,Dercks:2016npn}, and the
subsequently calculated $\chi^2$ 
from event counts in the signal regions of a large number of LHC searches. The $\chi^2$-estimate has been used as an input to train a neural
network based regression; the resulting neural network then provides a very fast
and reliable evaluation of the LHC search results, which can be used as an
input to global fits.

Neural networks allow predictions of complex data by overlapping
non-linear functions in an efficient manner, and are straightforward to
implement using open source libraries like
Tensorflow~\cite{tensorflow2015-whitepaper}, Theano~\cite{theano} and Keras~\cite{keras}.
Previous studies~\cite{Buckley:2011kc,Bornhauser:2013aya,Caron:2016hib}  have 
used neural
networks (and other machine learning techniques) mainly as a classifier between
allowed and excluded models. Going beyond this simple classification,
recently Gaussian processes~\cite{Rasmussen2004} were used to predict
the number of signal events for a simple SUSY
model~\cite{Bertone:2016mdy}. Here, we have applied neural networks as a
regression tool to predict $\chi^2$ values for an
ensemble of many signal regions. Compared to other regression methods,
neural networks excel when dealing with sparse sample sets due to
their non-linear nature. They can thus be valuable tools for predicting
LHC $\chi^2$ values of complex  BSM models.

In order to train and validate the neural network regression we have considered the
pMSSM-11 as an example of a complex and phenomenologically highly 
relevant SUSY model. Two different approaches have been explored: in
the 
so-called direct approach we have simply used the
eleven parameters of the pMSSM-11 as input to the regression, c.f.\ Figure~\ref{fig:nn_flow}a. 
The resulting neural network evaluates the $\chi^2$ of the pMSSM-11 within milliseconds and can thus be used in global 
pMSSM-11 fits without time penalty. A second, so-called
reparametrized, approach uses SUSY particle masses, cross sections and
branching ratios to first estimate signature-based objects, such as
particle energies and particle
multiplicities, which should relate more closely to the LHC $\chi^2$
values (Figure~\ref{fig:nn_flow}b). While the calculation of the
particle energies and 
multiplicities requires extra computation time, the corresponding neural
network should be more general and be able to predict the $\chi^2$ for
a wider class of SUSY models. To explore this feature, we have examined
how well a reparametrized neural network trained on the pMSSM-11 can predict the 
$\chi^2$ for two other popular SUSY models, namely the cMSSM and 
a model with anomaly mediated SUSY breaking
(AMSB)~\cite{Randall:1998uk,Giudice:1998xp}.

This paper is structured as follows: In section
\ref{sec:Event_generation_and_LHC_chi2_calculation} we provide details
of the LHC event generation within the pMSSM-11, and the method we have used to calculate a
global LHC $\chi^2$. The training and validation of the direct and
reparametrized neural net approaches are presented in 
\ref{sec:direct} and \ref{sec:reparameterized}, respectively. The results are
summarized and our conclusions presented in section~\ref{sec:conclusion}. More details
on the statistical approach are given in the appendices.

\section{Event generation and LHC $\chi^2$ calculation}\label{sec:Event_generation_and_LHC_chi2_calculation}
 
\begin{figure}[t]
\large
\begin{tikzpicture}[%
    >=triangle 60,              
    start chain=going below,    
    node distance=16mm and 36.5mm, 
    every join/.style={norm},   
    scale=0.8,
    every node/.style={transform shape},
    ]
\tikzset{
  base/.style={draw, on chain, on grid, align=center, minimum height=4ex},
  proc/.style={base, rectangle, text width=6em},
  proc2/.style={base, rectangle, text width=8em},
}
\node [proc] (p0) {pMSSM-11 parameters};
\node [proc, below=of p0] (p1) {SPheno};
\node [proc, below=of p1] (p2) {Madgraph};
\node [proc, below=of p2] (p3) {Pythia};
\node [proc, below=of p3] (p4) {CheckMATE};
\node [proc, below=of p4] (p5) {$\chi^2_{\rm CM}$};
\node [proc, below left=0mm and 33.7mm of p2] (p6) {NLL-Fast};

\node [proc2, right=of p1] (b1) {Spectrum calculation};
\node [proc2, right=of p2] (b2) {MC-Event generation};
\node [proc2, right=of p3] (b3) {Showering and hadronization};
\node [proc2, right=of p4] (b4) {Detector Simulation/implementation of LHC analyses};

\draw[->] (p0.south) --  (p1.north);
\draw[->] (p1.south) -- (p2.north);
\draw[->] (p2.south) -- (p3.north);
\draw[->] (p3.south) -- (p4.north);
\draw[->] (p4.south) -- (p5.north);
\draw[->] (p1.west) -| (p6.north);
\draw[-] (p6.east) -- (p2.west);
\draw[-] ($(p6)!.5!(p2)$) |- ($(p3)!.5!(p4)$) node[midway,above left=8mm and 0mm] {Cross sections};

\draw[dotted,line width=0.75pt] (p1.east) -- (b1.west);
\draw[dotted,line width=0.75pt] (p2.east) -- (b2.west);
\draw[dotted,line width=0.75pt] (p3.east) -- (b3.west);
\draw[dotted,line width=0.75pt] (p4.east) -- (b4.west);
\end{tikzpicture}
\caption{\sl Flow chart of the simulation chain we have used in the sample
  generation. The subscript ``CM'' in $\chi^2_{\rm CM}$ denotes $\chi^2$-values 
obtained using CheckMATE.}
\label{fig:simulation_chain} 
\end{figure}
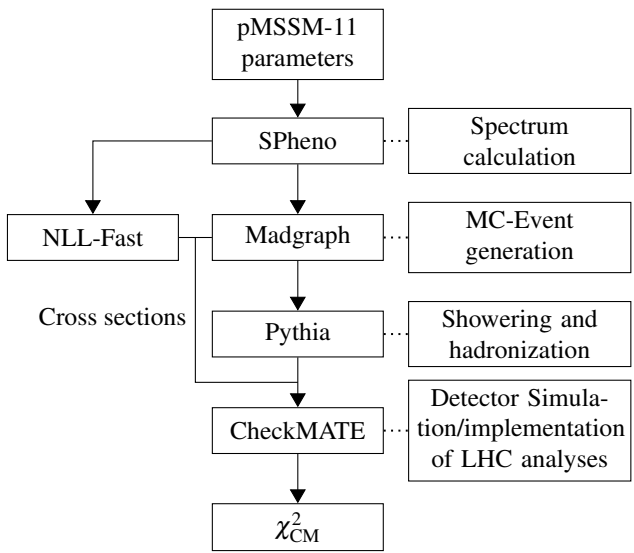

For the training and testing of the neural networks we have calculated the
LHC $\chi^2$ for a set of pMSSM-11 parameter points using CheckMATE. 
This $\chi^2_{\rm CM}$ calculation (the subscript ``CM'' denotes the
CheckMATE result) 
required the generation of SUSY signal events, a detector simulation 
and the evaluation of the LHC analyses. In this section we describe the simulation
chain, list the LHC ana\-ly\-ses we include, and explain our calculation of 
$\chi^2_{\rm CM}$ from the event counts in the various experimental signal regions. A graphical
depiction of this process can be found in
Figure~\ref{fig:simulation_chain}. 

\begin{table*}[t]
\centering
\begin{tabular*}{0.7\textwidth}{@{\extracolsep{\fill}}cll@{}} \hline
pMSSM-11 Parameter & Description & Range                        \\ \hline
$M_1$ & Bino mass               &  [-4000,4000] GeV               \\
$M_2$ & Wino mass               & [100,4000] GeV                 \\ 
$M_3$ & Gluino Mass               & [-4000,-400]$\cup$[400,4000]  GeV             \\ 
$M(\tilde{Q}_{12})$ & 1st and 2nd gen. scalar squark mass           & [300,5000] GeV                 \\ 
$M(\tilde{Q}_3)$ & 3rd gen. scalar squark mass           & [100,5000] GeV                 \\ 
$M(\tilde{L}_{12})$ & 1st and 2nd gen. scalar slepton mass           & [100,3000] GeV                 \\ 
$M(\tilde{L}_3)$ & 3rd gen. scalar slepton mass           & [100,4000] GeV                 \\ 
$M(A)$ & Pseudoscalar Higgs pole mass             & [0,4000] GeV                   \\ 
$A$ & Third generation trilinear coupling                & [-5000,5000] GeV \\ 
$\mu$ & Higgsino mass parameter              & [-5000,-100]$\cup$[100,5000] GeV             \\ 
$\tan(\beta)$ & Higgs doublet vacuum expectation value        & [1,60] GeV \\           
\end{tabular*}
\caption{\sl pMSSM-11 ranges we have used for the sample generation. The lower bounds on the mass parameters
were motivated by the current collider bounds for SUSY particle production. The 
relatively large range for third generation trilinear coupling $A$ were motivated by the observed
Higgs mass. All the pMSSM-11 parameters were defined at a scale of 1 TeV \cite{AguilarSaavedra:2005pw}.}
\label{pMSSM_11_parameters}
\end{table*}

The pMSSM-11 is based on eleven SUSY parameters: the bino, wino and gluino mass
parameters, $M_1,M_2, M_3$, the scalar quark and lepton masses of the 1st/2nd
and 3rd generation, respectively, $M(\tilde{Q}_{12})$,
$M(\tilde{Q}_{3})$, $M(\tilde{L}_{12})$,
$M(\tilde{L}_{3})$,  the mass of the pseudoscalar Higgs, $M(A)$, the
trilinear coupling of the 3rd generation, $A$, the Higgsino mass
parameter, $\mu$, and the ratio of the two vacuum expectation values,
$\tan(\beta)$~\cite{Djouadi:1998di}. The ranges of the pMSSM-11
parameters we have considered are specified in Table~\ref{pMSSM_11_parameters}.

We have sampled pMSSM-11 points using both a  uniform and a Gaussian random probability
distribution. The maximum and standard deviation of the Gaussian distribution have been chosen to be one fourth
  and one half of the parameter range, respectively.\footnote{For ranges 
  with allowed negative values, this distribution is mirrored around 
  zero.} Consequently we have generated less points near the edges of the parameter
range, and in the decoupling regime of large pMSSM-11 parameters, where
the SUSY spectra are beyond the LHC reach.

After we have generated a pMSSM-11 parameter point we have used SPheno-3.3.8
\cite{spheno} to calculate the spectrum. Following the SPA convention
\cite{AguilarSaavedra:2005pw}, all 11 parameters were
defined at a scale of 1 TeV. We have then only proceeded in
the simulation chain if the following preselection criteria have been
fulfilled:
\begin{itemize}
\item There were no tachyons in the spectrum.
\item The $\chi^0_1$ was the LSP, such that it was the dark matter candidate.
\item Both Higgs bosons ($h^0$, $H^0$) had a mass above 110 GeV. 
\item $m_{\chi_1^{\pm}}>$103.5 GeV \cite{Agashe:2014kda}  
\item The experimental value and the predicted value for the
  electroweak precision observables
  $ m_W,~ |\Delta (M_{Bs}),~ \mbox{BR}(B_s \rightarrow \mu \mu),~
  \mbox{BR}(b\rightarrow s \gamma),~ \mbox{BR}(B_u \rightarrow \tau
  \nu)$ differed by less than 5 times the total uncertainty given in
  Table \ref{table:LEO}.
\end{itemize}

The preselection criteria were applied in order to restrict the
SUSY parameter space to phenomenologically viable
regions. Consequently, the neural networks that were trained on that parameter
space will only be
valid if the preselection criteria are fulfilled. Note that the
anomalous magnetic moment of the muon, $(g-2)_{\mu}$, has not been included in the
preselection criteria, because in a global fit one might not wish to include
this observable. 

\begin{table*}[t]
\centering
\begin{tabular*}{0.7\textwidth}{@{\extracolsep{\fill}}llcc@{}}
\hline
Precision observable                     & Experimental value                       & Theoretical uncertainty    & Ref. \\ \hline
$m_W$                                      & $(80385 \pm 15) \mbox{MeV}$               &    0.1\%                & \cite{Group:2012gb}   \\
BR(b $\rightarrow$ s $\gamma$)           & $(3.43 \pm 0.21 \pm 0.07)\cdot10^{-4}$   & 14 \%                      & \cite{Amhis:2012bh}  \\
BR($B_s$ $\rightarrow$ $\mu^+$ $\mu^-$)  &  $(2.9 \pm 0.7)\cdot10^{-9}$           &  26 \%                     & \cite{CMSandLHCbCollaborations:2013pla}  \\
BR($B_u$ $\rightarrow$ $\tau$ $\nu$)     & $(1.05 \pm 0.25) \cdot 10^{-4}$           & 20 \%                     & \cite{Beringer:1900zz}  \\
$|\Delta(M_{Bs})|$                       & $(17.719 \pm 0.036 \pm 0.023)\ \mbox{ps}^{-1}$ & 24 \%                & \cite{Beringer:1900zz}    \\
\end{tabular*}
\caption{\sl Precision observables used to constrain the parameter space. Note that a newer and slightly more precise measurement BR($B_s$ $\rightarrow$ $\mu^+$ $\mu^-) = (2.8 ^{+ 0.7}_{-0.6})\cdot10^{-9}$ is now available but was not included in the preselection \cite{CMS:2014xfa}.}  
\label{table:LEO}
\end{table*}

\begin{table*}[t]
\centering
\begin{tabular*}{0.85\textwidth}{@{\extracolsep{\fill}}clllll@{}}
\hline
Nr. & Name & Description & $\mathcal{L}[$fb$^{-1}]$ & $N_{\rm SR}$ & Ref.                            \\ \hline 
\multicolumn{3}{l}{Analyses~@~8~TeV}  & &                        \\ \hline
1 & atlas\_1402\_7029 & 3 leptons + $\slashed{E}_{T}$ & 20.3 & 24 &\cite{ATLAS-1402-7029}               \\ 
2 & atlas\_1403\_5294 &2 leptons + $\slashed{E}_{T}$ & 20.3 & 9 &\cite{ATLAS-1403-5294}                \\ 
3 & atlas\_conf\_2013\_036 &4 or more leptons & 20.7 & 5 &\cite{ATLAS-CONF-2013-036}            \\ 
4 & atlas\_1308\_2631 &2 b-jets + $\slashed{E}_{T}$ & 20.1 & 6 &\cite{ATLAS-1308-2631}                 \\ 
5 & atlas\_1403\_4853 &2 leptons(opposite sign) & 20.3 & 12 &\cite{ATLAS-1403-4853}                  \\ 
6 & atlas\_1404\_2500 &2 leptons(same sign) + jets & 20.3 & 5 &\cite{ATLAS-1404-2500}                  \\ 
7 & atlas\_1405\_7875 &Multiple jets + $\slashed{E}_{T}$ & 20.3 & 15 &\cite{ATLAS-1405-7875}                 \\ 
8 & atlas\_1407\_0583 &1 isolated lepton + jets + $\slashed{E}_{T}$& 20.3 & 27 &\cite{ATLAS-1407-0583}                   \\ 
9 & atlas\_1407\_0608 &2 stop quarks& 20.3 & 3 &\cite{ATLAS-1407-0608} \\ 
10 & atlas\_1502\_01518 &1 jet + $\slashed{E}_{T}$& 20.3 & 9 &\cite{ATLAS-1502-01518}             \\ 
11 & atlas\_1503\_03290 &2 leptons(same flavor, opposite sign) + $\slashed{E}_{T}$& 20.3 & 1 &\cite{ATLAS-1503-03290}                     \\ \hline
\multicolumn{3}{l}{Analyses~@~13~TeV} & &                           \\ \hline 
12 & atlas\_conf\_2015\_076 &1 isolated lepton + jets + $\slashed{E}_{T}$ & 13.3 & 6 &\cite{ATLAS-CONF-2015-076}                \\ 
13 & atlas\_1602\_09058 &Jets + 2 leptons(same sign) or 3 leptons& 3.2 & 4 &\cite{ATLAS-1602-09058}                \\ 
14 & atlas\_1605\_03814 &Multiple jets + $\slashed{E}_{T}$& 3.2 & 7 &\cite{ATLAS-1605-03814}               \\ 
15 & atlas\_conf\_2015\_082 &Leptonically decaying Z + jets + $\slashed{E}_{T}$& 3.2 & 1 &\cite{ATLAS-CONF-2015-082}                 \\ 
16 & atlas\_1604\_07773 &1 jet + $\slashed{E}_{T}$& 3.2 & 13 &\cite{ATLAS-1604-07773}                \\  
17 & atlas\_conf\_2016\_013 &Multiple leptons + Multiple Jets & 3.2 & 10 &\cite{ATLAS-CONF-2016-013}                \\ 
18 & atlas\_conf\_2015\_067 &b-jets + $\slashed{E}_{T}$& 3.2 & 3 &\cite{ATLAS-CONF-2015-067}                   \\ 
19 & cms\_pas\_sus\_15\_011 &2 leptons(same flavor, opposite sign) + $\slashed{E}_{T}$& 2.2 & 47 &\cite{CMS-PAS-SUS-15-011}                  \\
\end{tabular*}
\caption{\sl Analyses considered at 8~TeV and 13~TeV. The names
  correspond to the internal  CheckMATE nomenclature. $\mathcal{L}$
  denotes the luminosity and $N_{\rm SR}$ the number of SRs for each analysis.
}
\label{table:analyses}
\end{table*}

\subsection{Event generation}\label{sec:event_generation}

If all pre-selection criteria were fulfilled we then generated LHC
signal events for 8 and 13~TeV. At 8~TeV we simulated the production of 
both electroweak and strongly interacting SUSY particles, including all
possible $2\to2$ scattering processes. The 8~TeV analyses were 
included since no 13~TeV electroweak searches were available 
at the time the event generation for this work was started. For the simulation of the
electroweak events, which include all combinations of slepton, chargino
and neutralino final states, we used MadGraph5\_aMC@NLO
2.4 \cite{madgraph} with the CTEQ6L1 PDF \cite{Nadolsky:2008zw}
and showered the events with Pythia 6.428 \cite{pythia6_4_manual}.
Since the LHC is only sensitive
to electroweak production if
$m_{\chi^0_1} < 500$~GeV~\cite{sensitivity_8TeV_ATLAS_EW}, we have only simulated 
the electroweak processes if this was the case. 
The processes which contain final states with strongly interacting SUSY
particles, \textit{i.e.} squarks and gluinos, 
were simulated using Pythia 8.2 \cite{pythia8,Desai:2011su}, with cross
sections normalized to NLO using NLL-Fast~2.1~\cite{nll_fast_1,nll_fast_2,nll_fast_3,nll_fast_4,nll_fast_5,nll_fast_6,nll_fast_7,nll_fast_8}.

\begin{figure}[t] 
  \begin{minipage}{\columnwidth}
    \centering
    \includegraphics[width=\textwidth]{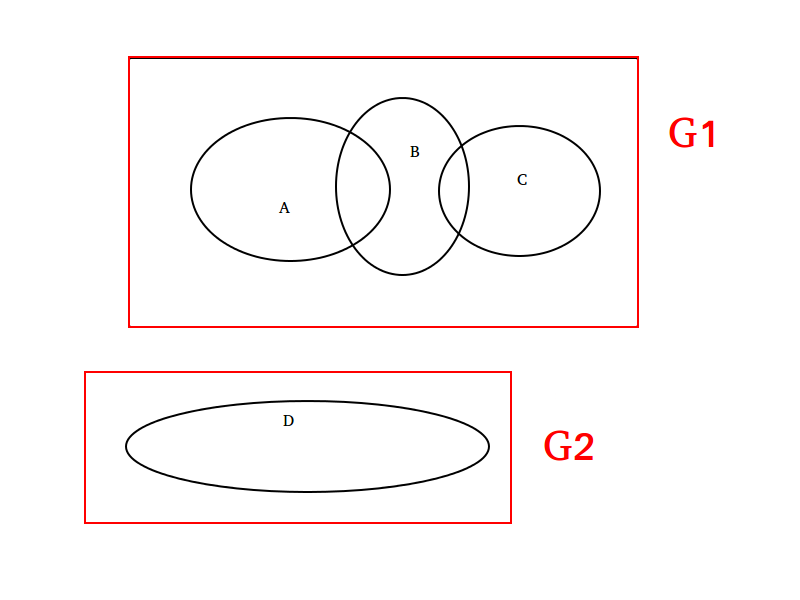}
  \end{minipage}
  \caption{\sl Example for a disjoint group selection for one
    analysis. The analysis has four SRs (A,B,C,D). Because SR A
    overlaps with SR B and SR B overlaps with SR C, we grouped SR A,B and
    C into one disjoint group G1. Because SR D does not overlap with
    any of the other SR of the analysis, the disjoint group G2 only
    contains the SR D.}
 \label{fig:disjoint_groups}
\end{figure}

At 13~TeV, only processes with strongly interacting SUSY particles were
simulated, since all analyses which were available within CheckMATE
target 
these final states. We have again used MadGraph5\_aMC@NLO~5.2.4 \cite{madgraph} to generate events and Pythia 6.248
\cite{pythia6_4_manual} was used to decay and shower the final state. The cross
sections were normalized to NLO accuracy as obtained from
NLL-Fast~2.1~\cite{nll_fast_1,nll_fast_2,nll_fast_3,nll_fast_4,nll_fast_5,nll_fast_6,nll_fast_7,nll_fast_8}. One
additional
parton was generated at the matrix element level and then matched to
the parton shower if the mass gap between either the lightest squark
or the gluino and the LSP was below 300 GeV.  Such a procedure allowed
for the accurate determination of the signal acceptance in compressed
spectra where initial state radiation is crucial to pass the
experimental cuts \cite{Dreiner:2012gx,Dreiner:2012sh}.

\begin{figure*}[t]
  \begin{subfigure}[b]{0.5\textwidth}
    \includegraphics[width=\textwidth]{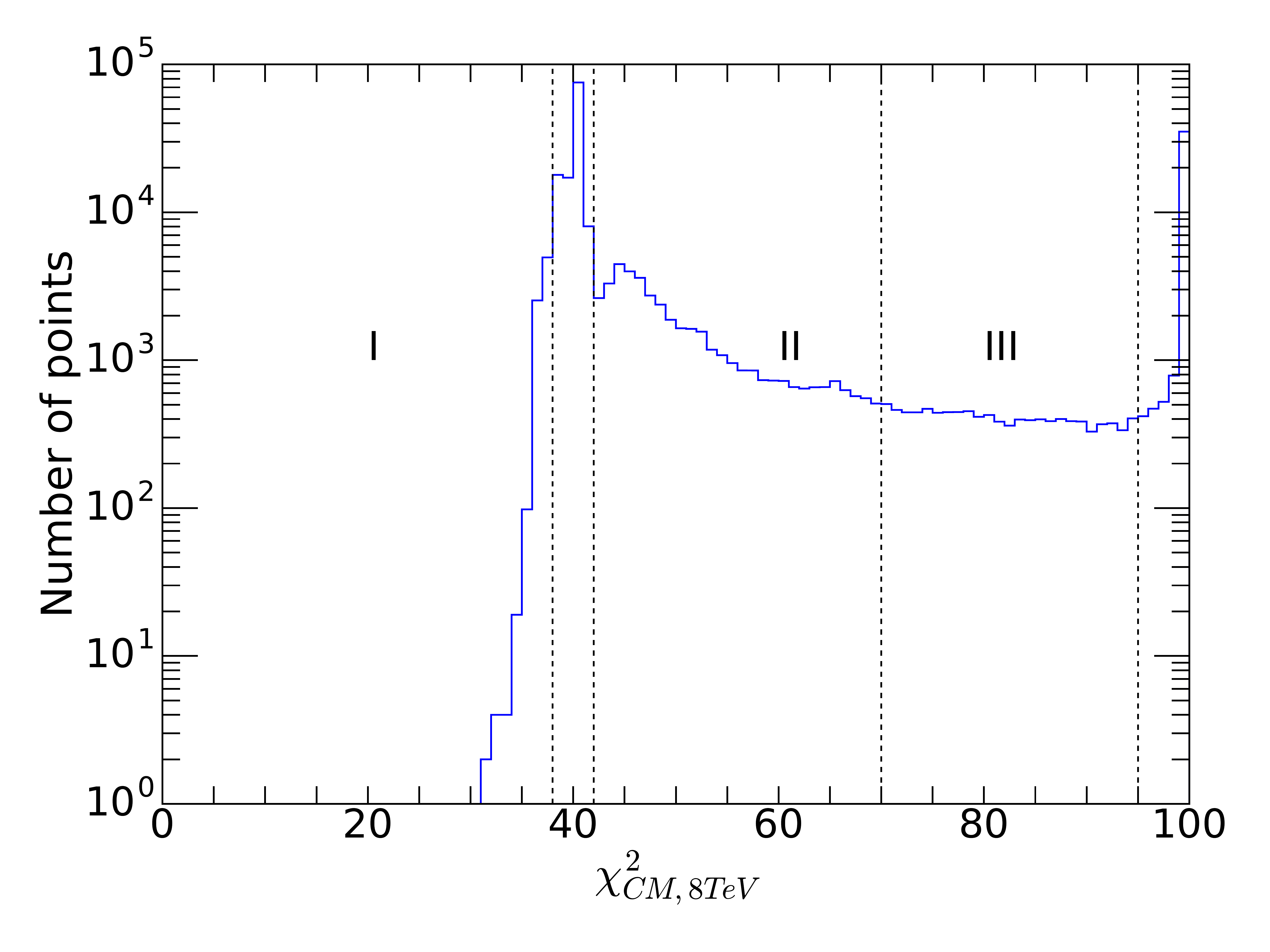}
    \caption{}
    \label{fig:chi2_distribution_8TeV}
  \end{subfigure}
  \hfill
  \begin{subfigure}[b]{0.5\textwidth}
     \includegraphics[width=\textwidth]{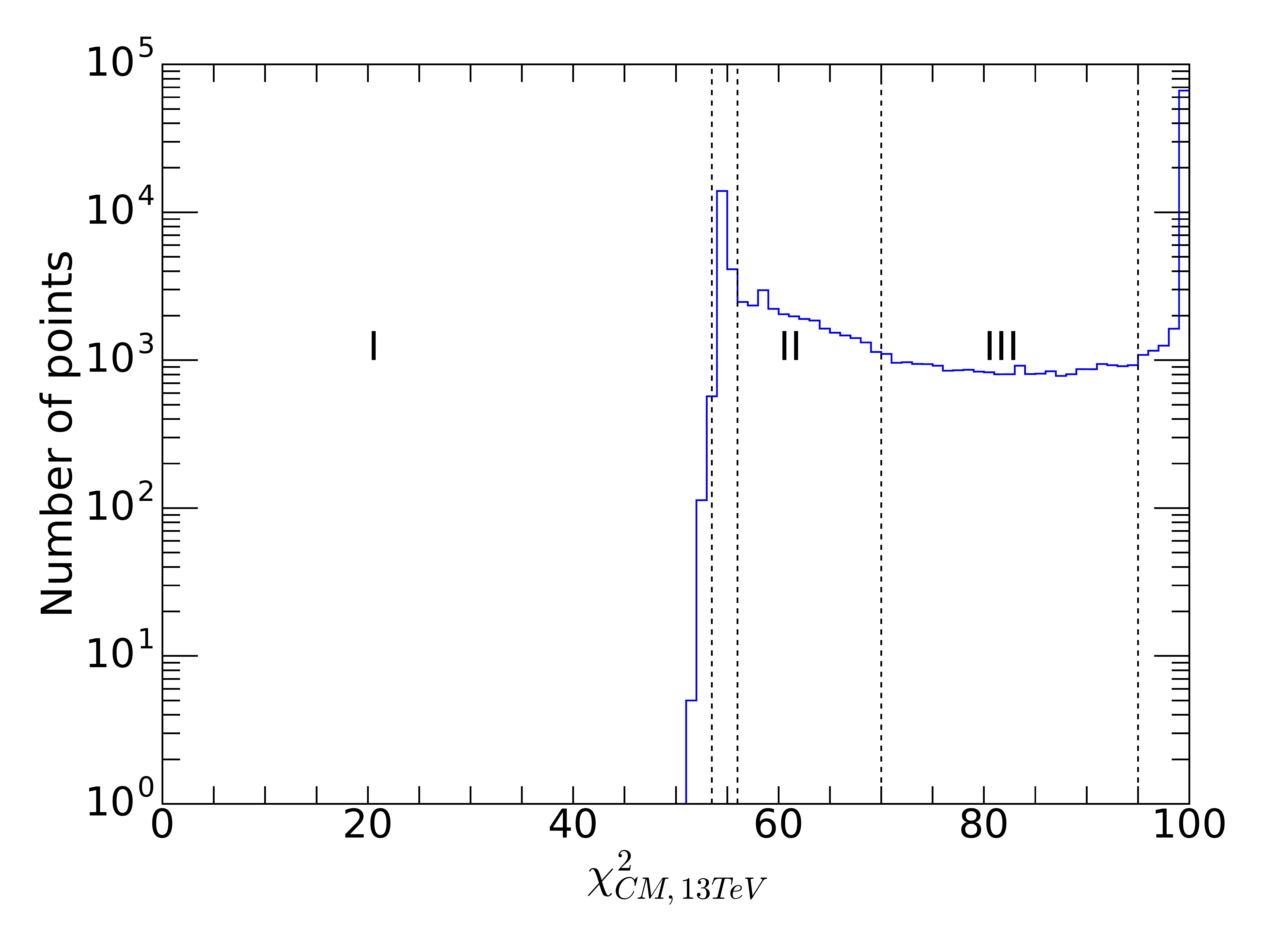} 
    \caption{}
    \label{fig:chi2_distribution_13TeV}
  \end{subfigure}
  \caption{\sl Distributions of $\chi^2_{\rm CM}$ for 8~TeV (a) and
    13~TeV (b). The dashed lines indicate ranges for which we evaluated
    the performance of the SCYNet neural network separately as
    described in the text.}
\end{figure*}

For 8\,TeV we calculated $\chi^2_{\rm CM}$ for a total of 210000 pMSSM-11
parameter points and at 13\,TeV
we simulated 140000 parameter points.
To guarantee a sufficient statistical accuracy, we have generated between
1000 and 45000 events per particle production process, depending
in detail on the expected number of signal events. For each pMSSM-11
parameter point we have simulated on average 20000 events, which required
approximately 380000 CPU hours in total.

Once the event
generation was completed, the event files were passed through
CheckMATE\footnote{CheckMATE uses the FastJet library
  \cite{Cacciari:2005hq,Cacciari:2011ma} and in particular the anti-kt
  algorithm \cite{Cacciari:2008gp} for jet reconstruction. In the course of this
  work we have used analyses originally developed for SUSY searches in the NMSSM \cite{Cao:2015ara}
  and the Super-Razor observable \cite{Buckley:2013kua}.
  }
\cite{Drees:2013wra,Kim:2015wza,Dercks:2016npn} which contains a tuned
Delphes-3 \cite{delphes} detector simulation with separate setups for
8 and 13~TeV. The analyses which have been used to train the SCYNet neural
network regression are listed in Table~\ref{table:analyses}.

\subsection{LHC $\chi^2$ calculation with CheckMATE}\label{subsection:LHC_chi2_calculation}
We have implemented a calculation of $\chi^2_{\rm CM}$ that approximates to a
likelihood, based on the CheckMATE output of the event count in
each signal region (SR). In the
following we will use $\chi^2_{{\rm CM},jk}$ to denote
the $\chi^2$ of analysis
$j$ and SR $k$ from CheckMATE. The exact statistical prescription we have used to
calculate $\chi^2_{{\rm CM},jk}$ for each {\it single} SR is described in
\ref{chi2_SR}.  

The individual signal regions were combined to give a global
$\chi^2_{\rm CM}$ with a procedure
that combined the most sensitive (expected) orthogonal SRs. Our algorithm
chose these by first dividing the SRs in each analysis
into orthogonal disjoint groups. Each disjoint group can contain several SRs and the SRs
of one disjoint group are disjoint to all SRs in all other disjoint groups, see Figure \ref{fig:disjoint_groups}. 
If one disjoint group contains more than one SR, the SR with the largest
$\mbox{signal}/S^{95}_{\mbox{exp}}$ ratio in this group was
selected. For one analysis we then added the $\chi^2$s of all selected
SRs,
$  \chi^2_{{\rm CM},j}:=\sum\limits_{\mathclap{{\rm selected}\; k}}\
  \chi^2_{{\rm CM},jk}$.  
In the last step all the $\chi^2_{{\rm CM},j}$ from the individual analyses have been added to give
the overall
 $\chi^2_{\rm CM}:=\sum\limits_{j}\chi^2_j$. 
 All the analyses we have combined for the same centre-of-mass energy target
 different final state topologies, 
and we have therefore assumed that the SRs of different analyses were to a very good
approximation disjoint. In the case of the 8 (13) TeV analyses we found 47 (65) disjoint SRs 
for the group with the largest sensitivity.

Since the experimental collaborations do not provide information on
the correlation of the systematic errors, we had to assume the uncertainties were Gaussian distributed and
uncorrelated.

Figures~\ref{fig:chi2_distribution_8TeV} and
\ref{fig:chi2_distribution_13TeV} show the distributions of
$\chi^2_{\rm CM}$ for all sampled points. Two peaks can be 
observed in both plots: the first peak, called the \textit{zero
signal peak}, is located at around  $\chi^2_{\rm CM}\approx 40$ (8~TeV, left
plot) and $\chi^2_{\rm CM}\approx 54$ (13~TeV, right plot), respectively. 
All $\chi^2_{\rm CM}$ values in the zero signal
peaks belong to pMSSM-11 parameter points with very heavy SUSY spectra 
and thus very little or no expected signal in any of the SRs. 
A second peak occurs at $\chi^2=100$ because we set all $\chi^2_{\rm
  CM} > 100$ to this value. This was done to avoid the neural
networks learning any structures at large $\chi^2_{\rm CM}>100$,  
where the model is already clearly excluded by the LHC data. 

 The three regions labeled I, II and III outside of the zero
signal peak and the peak at $\chi^2_{\rm CM} = 100$ in Figures~\ref{fig:chi2_distribution_8TeV} and
\ref{fig:chi2_distribution_13TeV} are called \textit{rare target
  ranges}. We differentiated between regions below (I) and above (II and III) the zero
signal peak and between regions with $\chi^2_{\rm CM}$ closer to the
minimum (I and II), since these are much more important for
global fits than those far away, and thus need better modeling.
We have evaluated the performance of the SCYNet neural network separately
for these different regions.

\section{Neural network regression: direct approach}\label{sec:direct}
In this section the direct approach for neural network
regression used in SCYNet has been described. The inputs to the network were the 11 parameters of the 
pMSSM-11 and the output, $\chi^2_{\rm SN}$, is an estimate of $\chi^2_{\rm CM}$ for the LHC searches described in 
section~\ref{sec:Event_generation_and_LHC_chi2_calculation}. We have
trained separate networks for the 8 and 13~TeV analyses, with outputs
$\chi^2_{\rm SN, 8\,TeV}$ and $\chi^2_{\rm SN, 13\,TeV}$,
respectively. In this section, the setup and
performance of both networks has been discussed, but we have focused on the results obtained for
$\chi^2_{\rm SN, 8\,TeV}$ for illustration.

\begin{figure*}[t]
  \begin{subfigure}[b]{0.5\textwidth}
    \includegraphics[width=0.98\textwidth, height=0.283\textheight]{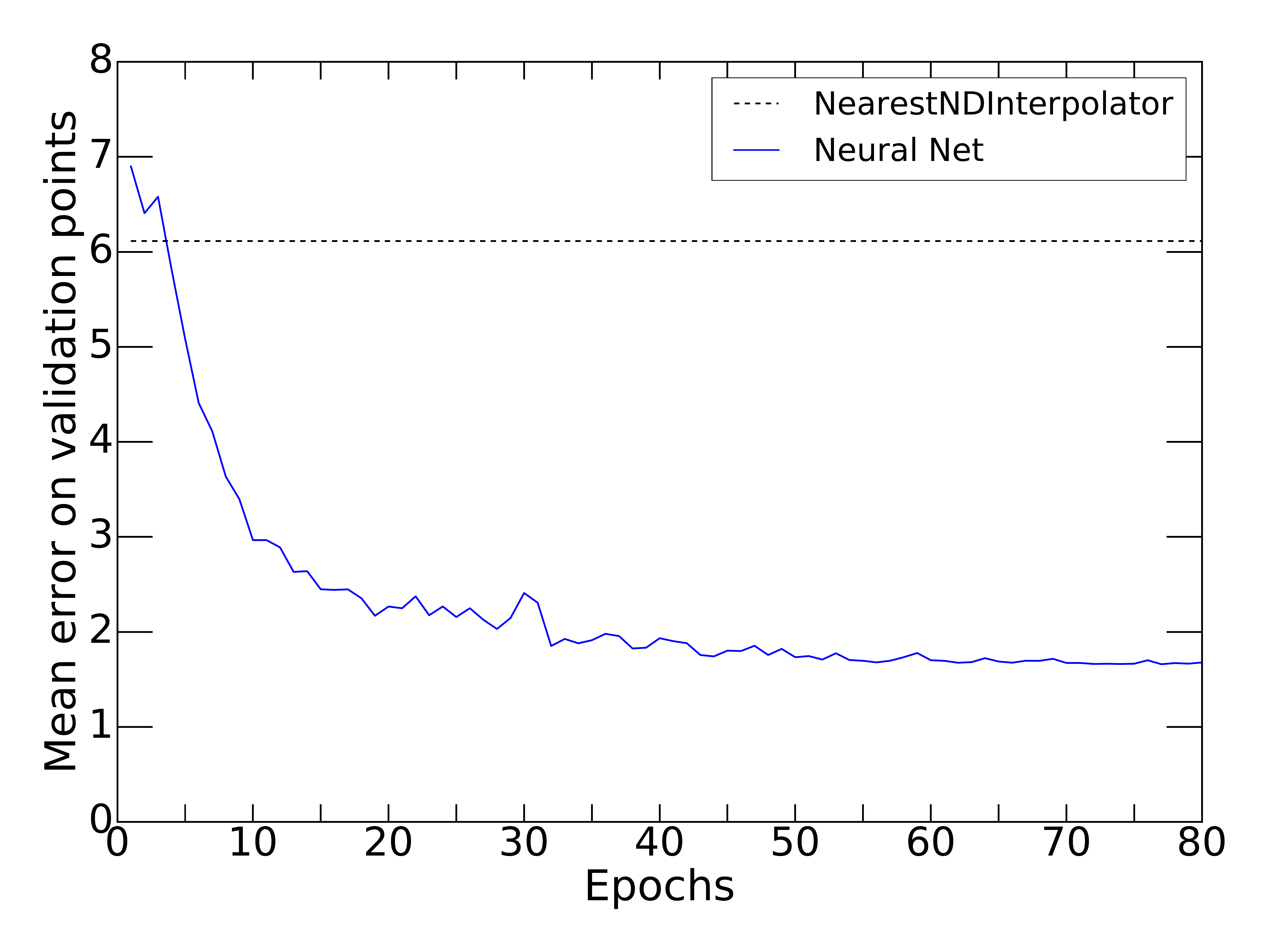}
    \caption{}
    \label{fig:training_8TeV_LHC_chi2_all}
  \end{subfigure}
  \hfill
  \begin{subfigure}[b]{0.5\textwidth}
     \includegraphics[width=\textwidth]{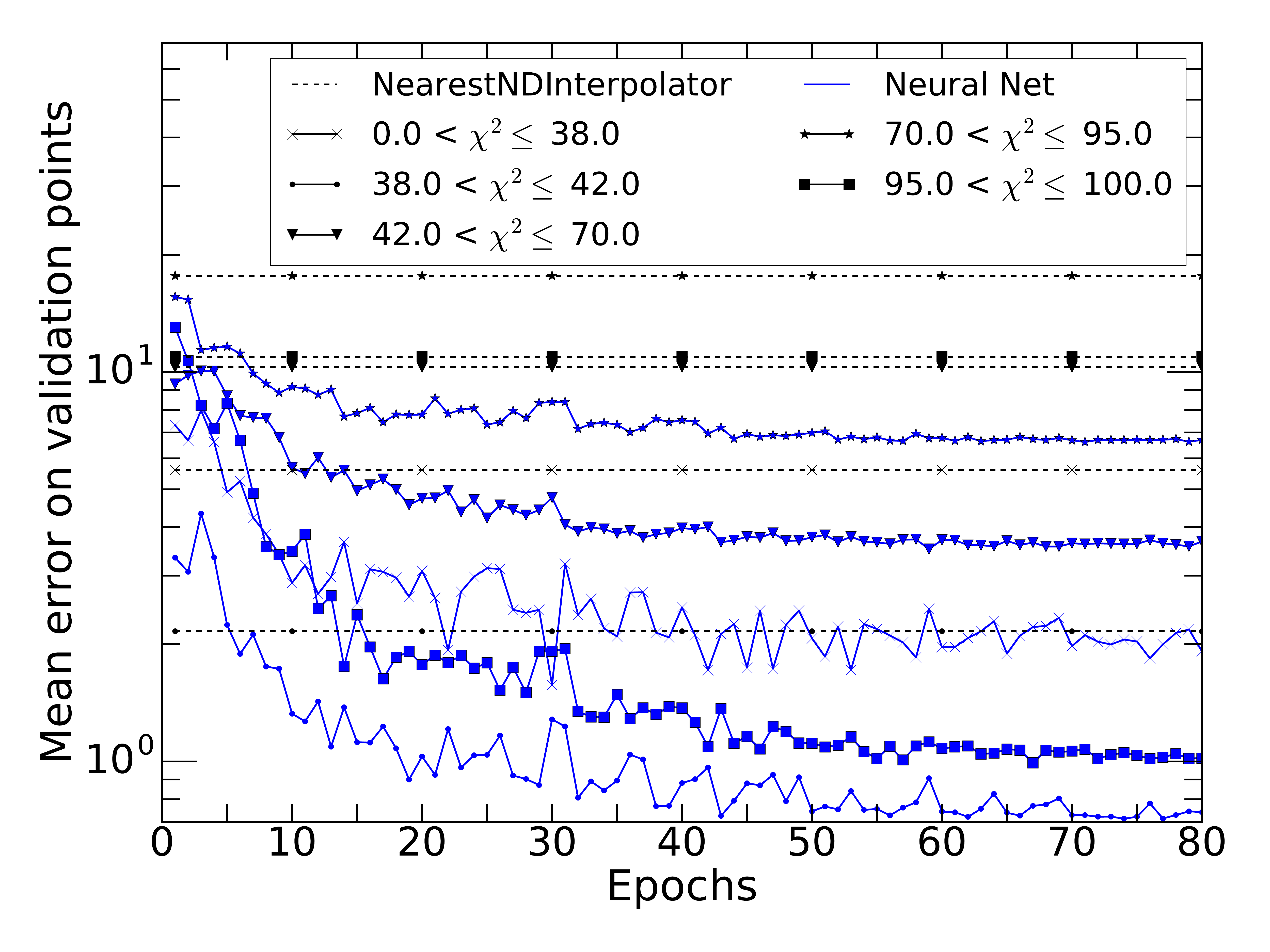}
    \caption{}
    \label{fig:training_8TeV_LHC_chi2_ranges}
  \end{subfigure}
      \caption{\sl Evolution of the mean error on
    points in the validation set during training on the 8~TeV LHC $\chi^2_{\rm CM}$ compared to a nearest neighbour interpolator 
    for all points (a), divided into different $\chi^2_{\rm CM}$ ranges (b).}
\end{figure*}

\subsection{Parameters of the neural networks}
\label{subsec:NN_param}

All neural networks used in this work were of the simple feed forward type\footnote{A feed 
forward neural network is `simple' if neurons from the $l$th layer are only connected with neurons in the $l+1$th layer.} and have been 
designed using the Tensorflow~\cite{tensorflow2015-whitepaper} library.
We performed scans of the neural network parameters (so-called
hyperparameters) to find the best configuration. The hyperparameters,
which we were trying to optimize, were the number of hidden layers, 
the number of neurons in the hidden layers,
the dropout probability of neurons in the hidden layers, the batch size, the learning rate of the Adam minimization 
algorithm~\cite{adam_optimizer}, the activation 
function in the output neuron, the regularization parameter and the type of cost function. The 
networks discussed in this section always refer to the hyperparameter configuration that was found 
to function best.\footnote{The optimal hyperparameter configuration is the one which produces 
the smallest average error between $\chi^2_{\rm SN}$ and $\chi^2_{\rm CM}$ 
on the validation set. The results of the hyperparameter optimization
are independent on the $\chi^2_{\rm CM}$ range considered. To prevent overfitting of the hyperparameters, for each scanned hyperparameter we choose 
a random validation set.} 

Our 8\,TeV network had four hidden layers $l=1,\ldots, 4$, each with $N_{l}=300$ neurons and all neurons 
have hyperbolic tangent activation functions. The weights in layer $l$ were initialized with a Gaussian distribution 
with standard deviation $1/\sqrt{N_{l-1}}$ and mean zero, while the
biases were initialized with a Gaussian distribution with standard
deviation equal to one and mean equal to zero. In 
order to train a network with a hyperbolic tangent output activation function, the targets were transformed to the range 
between $-1$ and $1$ with a modified Z-score normalization (see details in \ref{sec:modified_z_score_normalization}).

\begin{table*}[t]
\centering
\begin{tabular*}{0.8\textwidth}{@{\extracolsep{\fill}}lccccc@{}}
8 TeV \\\hline
$\chi^2$ range & $0- 38 $ & $38- 42 $& $42- 70 $ & $70- 95 $ &$95- 100 $ \\\hline
Direct &             1.9 &  0.7 &  3.7 &  6.7 &  1.0        \\
Reparameterized &   1.6 & 0.5 & 3.6 & 7.3 & 1.0 \\\hline  \\
13 TeV \\\hline
$\chi^2$ range & $0- 53.5 $ & $53.5- 56 $& $56- 70 $ & $70- 95 $ &$95- 100 $ \\\hline
Direct     &      1.8 &  0.9 &  2.4 &  4.0 &  0.5 \\
Reparameterized &   2.0 & 0.7 & 2.2 & 4.7 & 0.3 \\
\hline
\end{tabular*}
\caption{\sl Mean errors of the neural networks for the 8~TeV and 13~TeV LHC $\chi^2$ prediction in different ranges. The 
results for the direct and reparameterized approach (section~\ref{sec:reparameterized}) are compared.}
\label{NN_performance_ranges_8TeV}
\end{table*}

\begin{figure*}[t]
\begin{subfigure}[b]{0.5\textwidth}
    \includegraphics[width=\textwidth]{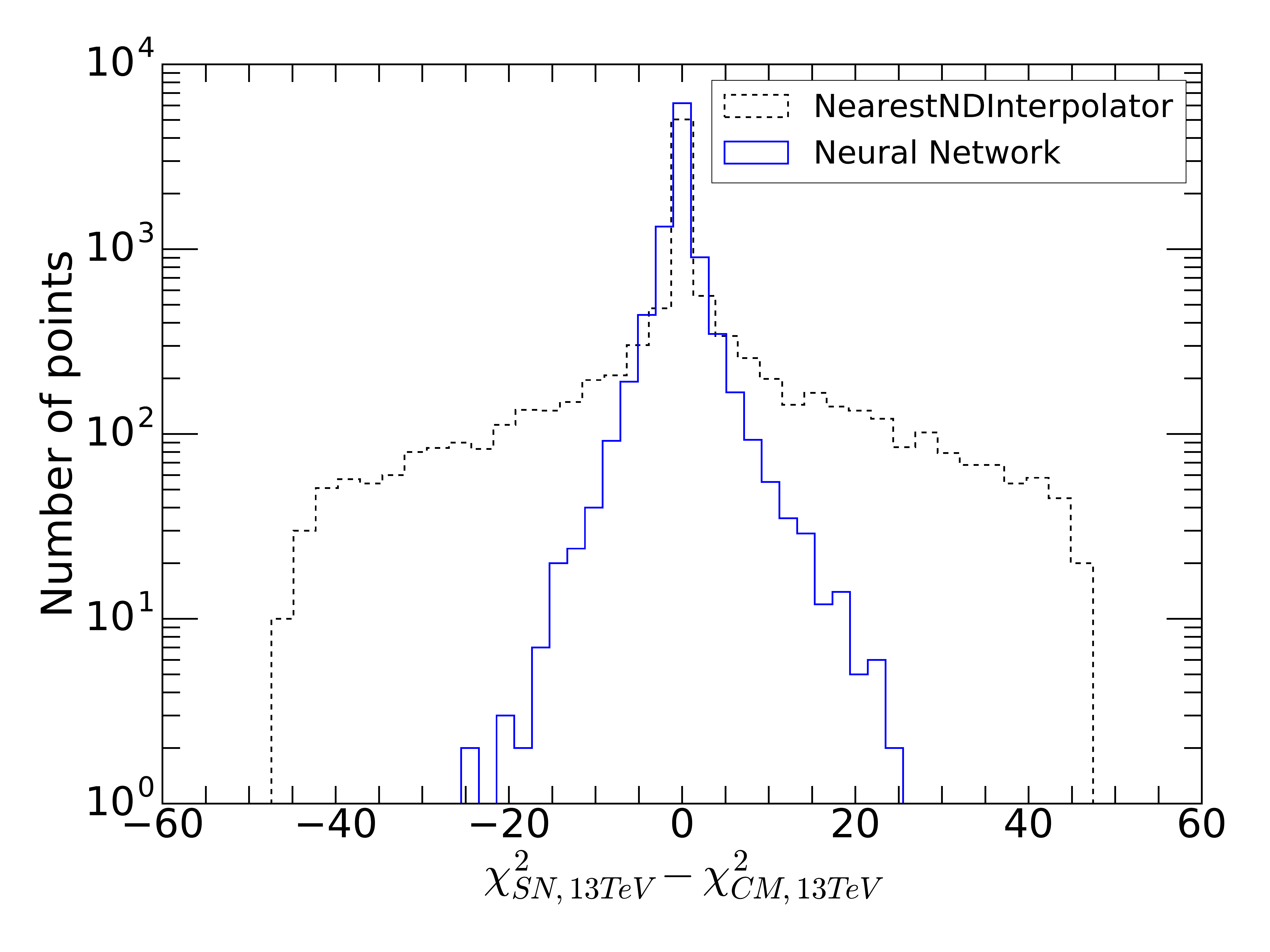}
    \caption{}
    \label{fig:13TeV_LHC_chi2_all}
  \end{subfigure}
  \hfill
  \begin{subfigure}[b]{0.5\textwidth}
     \includegraphics[width=\textwidth]{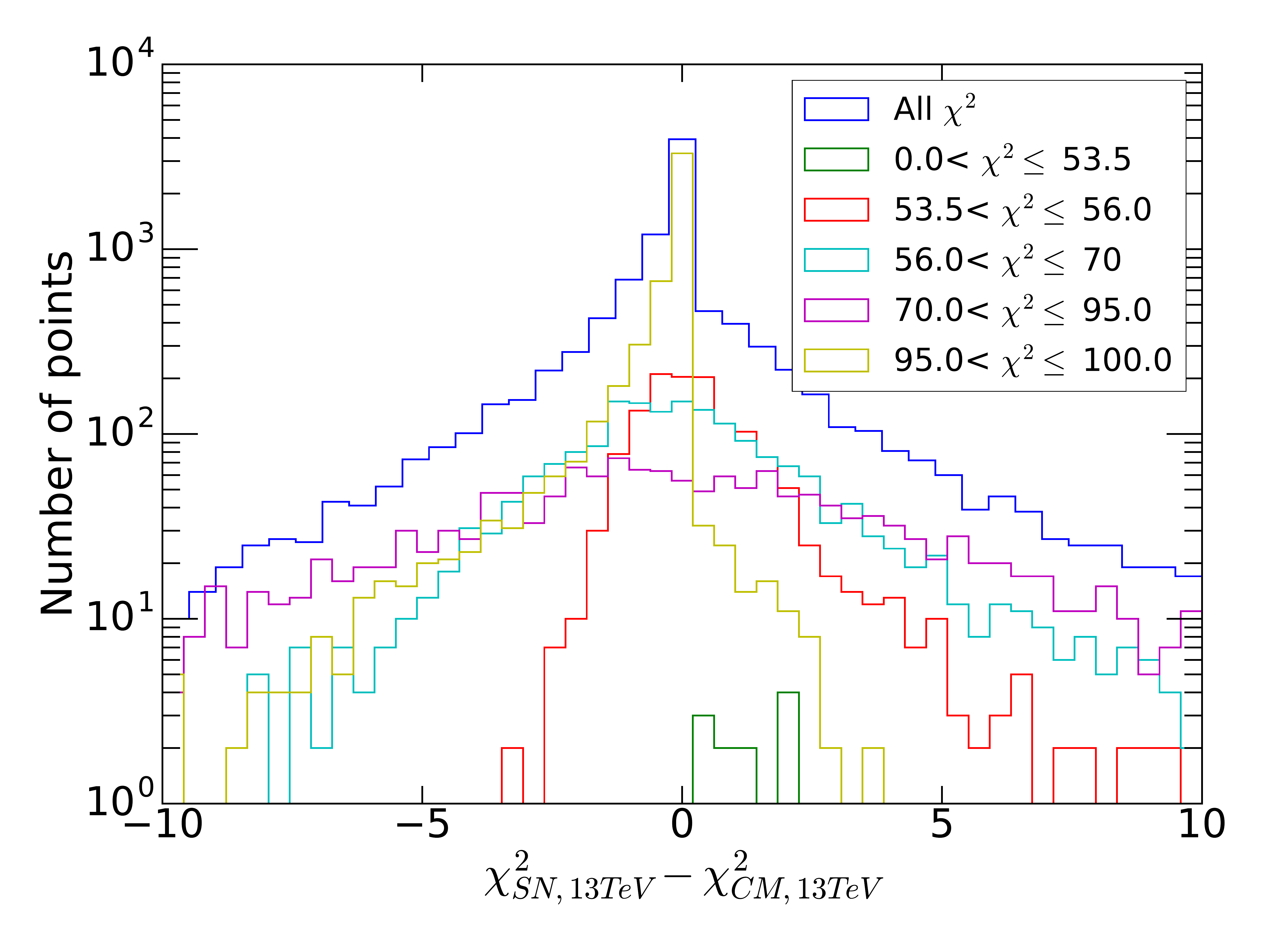}
    \caption{}
    \label{fig:13TeV_LHC_chi2_ranges}
  \end{subfigure}
  \caption{\sl Error histogram for the neural
    network compared to the 
  nearest neighbor interpolator for all validation points (a), divided by $\chi^2$ ranges (b). }
\end{figure*}

During the training we used a quadratic cost function and trained with a batch size of 750. We 
used the Adam minimization algorithm with a learning rate of 0.001 and all other parameters 
of the Adam optimizer were set to the default values from~\cite{adam_optimizer}.

The complete cost function $C$ consists of the quadratic cost function and a quadratic regularization:
\begin{equation}
C = \mbox{quadratic cost}+\frac{\lambda}{2 N_{{\rm train}}}\sum\limits_{jkl} (w_{jk}^l)^2, \label{eq:cost_func}
\end{equation}
where $N_{{\rm train}}$ were the number of points in the training set. We used 10000 validation 
points ($N_{{\rm val}}=10000$) while the rest of the sampled points were used for training.\footnote{No test 
set was used because in the hyperparameter scan the validation set was chosen randomly for each scanned hyperparameter.}
In the hyperparameter scan we found that $\lambda=10^{-5}$ gave the best network performance.

The 13\,TeV network had the same structure as the 8 TeV network but the batch size during training was slightly adjusted to
500, while all other hyperparameters were the same as in the 8\,TeV case.

\subsection{Training the neural networks}\label{sec:NN_training}

The training phase of the 8\,TeV neural network has been visualized in Figures~\ref{fig:training_8TeV_LHC_chi2_all}
and \ref{fig:training_8TeV_LHC_chi2_ranges} where we have compared the mean error on points 
in the validation data during the training phase to that of a nearest neighbor interpolator. The 
following parameter has been used to quantify the network performance,
\begin{equation}
\mbox{Mean error on valid. points} := \frac{1}{N_{{\rm
      val}}}\sum\limits_{i=1}^{N_{{\rm val}}} |\chi^2_{{\rm
    CM}_i}-\chi^2_{{\rm SN}_i}|.
\end{equation}
After the last training epoch the mean error on points in the validation set is 1.68. 
As shown in Figure~\ref{fig:training_8TeV_LHC_chi2_all} this performance is a significant improvement 
over a nearest neighbor interpolator~\cite{scipy}.

\begin{figure*}[t]
\begin{subfigure}[b]{0.5\textwidth}
    \includegraphics[width=\textwidth]{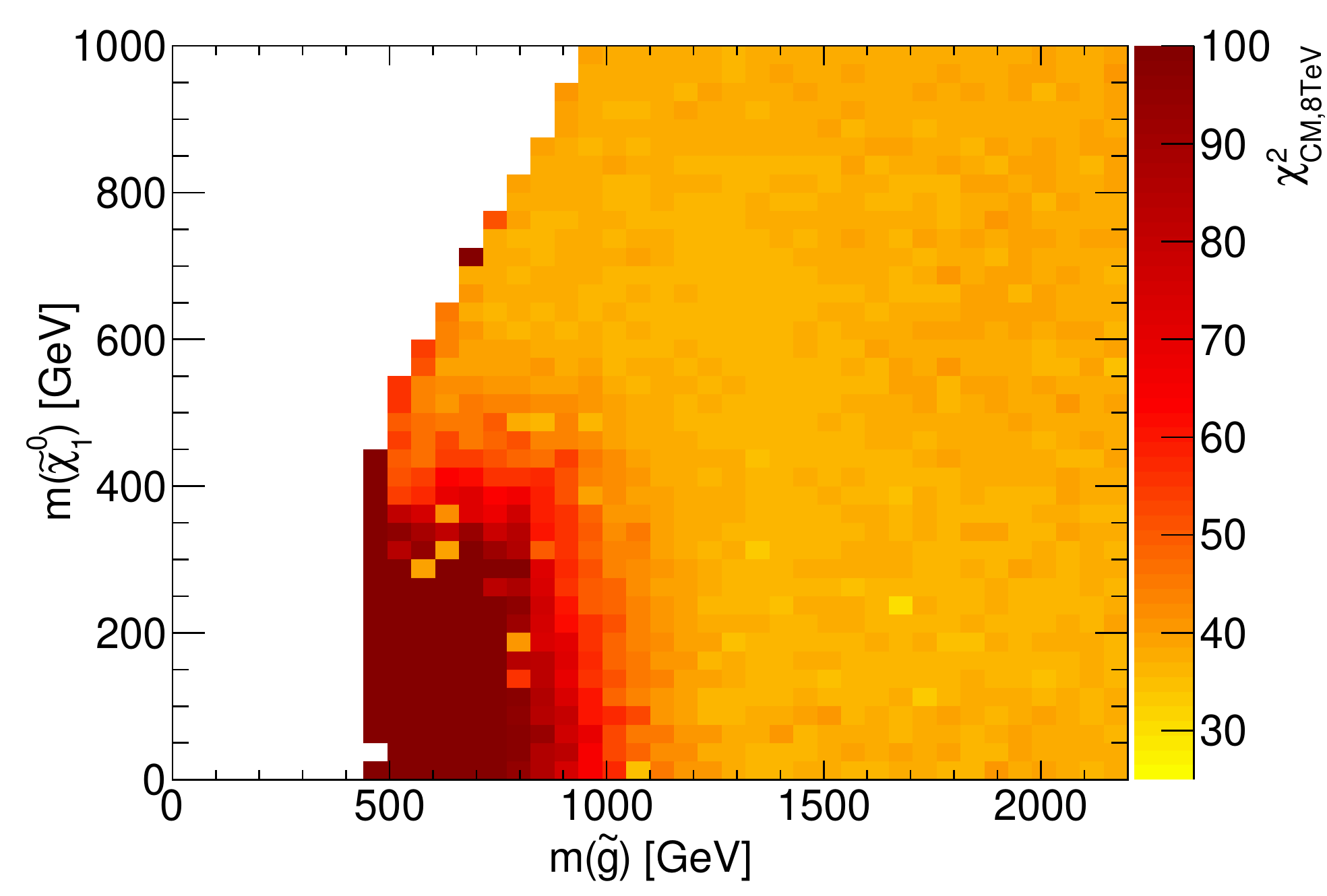}
    \caption{}
    \label{fig:profile_min_8TeV_mgo_mchi01_CM}
  \end{subfigure}
  \hfill
  \begin{subfigure}[b]{0.5\textwidth}
     \includegraphics[width=\textwidth]{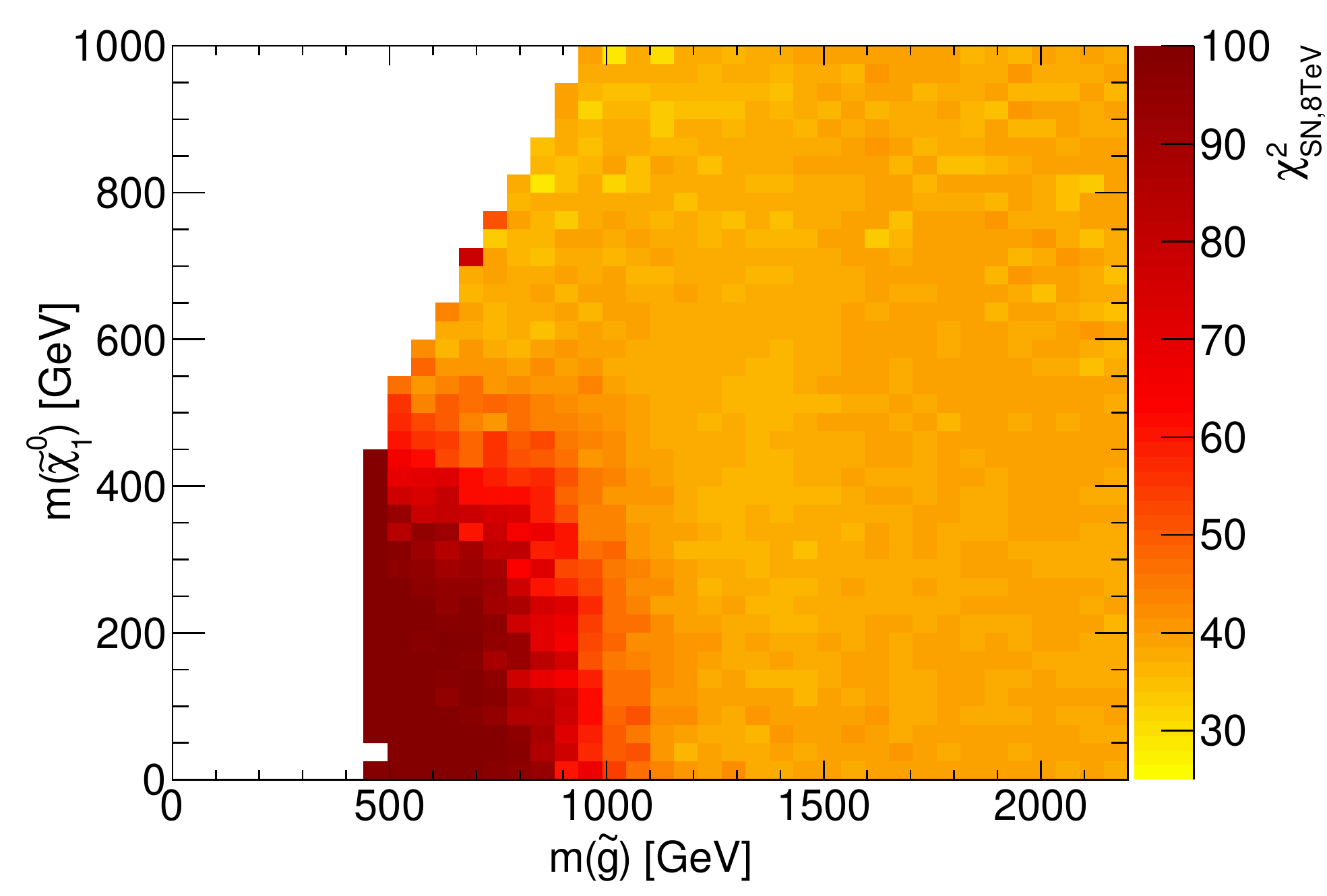}
    \caption{}
    \label{fig:profile_min_8TeV_mgo_mchi01_SN}
  \end{subfigure}
  \caption{\sl Minimum pMSSM-11 $\chi^2$ in the gluino-neutralino mass
    plane for the 8~TeV validation set for the CheckMATE
    results (a) and the predictions of SCYNet (b).}
\end{figure*}

Figure~\ref{fig:training_8TeV_LHC_chi2_ranges} shows how the mean
errors for $\chi^2$ values in the different ranges defined in
Figures~\ref{fig:chi2_distribution_8TeV} and \ref{fig:chi2_distribution_13TeV} 
behaved during the training phase. The corresponding mean errors after the last training epoch can be 
found in Table~\ref{NN_performance_ranges_8TeV}. The 
mean errors in the rare target ranges I ($0<\chi^2 \leq 38$), II
($42<\chi^2 \leq 70$), and III ($70<\chi^2 \leq 95$) were
larger than the mean errors in the zero signal range and in the range around 100. This behaviour will be called 
\textit{rare target learning problem} (RTLP) in the following and can
be understood from the $\chi^2_{\rm CM}$ distribution in Figure
\ref{fig:chi2_distribution_8TeV} and the corresponding 
discussion in 
section~\ref{subsection:LHC_chi2_calculation}: the majority of the scanned 
pMSSM-11 points led to $\chi^2_{\rm CM}$ values in the range
$38<\chi^2 \leq 42$ and $95<\chi^2$, which were thus described
more accurately by the neural network than those in regions I, II and
III. 

In future work the RTLP will be addressed with two strategies: the 11-dimensional probability density 
function (pdf) in the parameter space of the points in the range
$0<\chi^2 \leq 38$  and $42<\chi^2 \leq 95$
will be used to randomly sample new pMSSM-11 points which will be added to the training 
and validation samples. In the second approach, each point in the RTLP region will be used to seed 
new random points using a narrow 11-dimensional pdf centered around each 
point in the above-mentioned $\chi^2$ range. New points can then be
generated, simulated and used to improve the network especially in the
rare target ranges. Motivated by the profile likelihood requirement, which sets an estimate of $\Delta\chi^2=1$ for the $1\,\sigma$ range, 
we plan to improve the training and validation set size by subsequent application 
of these procedures until a mean error of $\Delta\chi^2$ well under 1 is
reached.

For the 13 TeV neural network we found similar results to those from the 8\,TeV network that has
already been displayed in Figure~\ref{fig:training_8TeV_LHC_chi2_all} and \ref{fig:training_8TeV_LHC_chi2_ranges}. 
The mean errors for the 13\,TeV network at the end of the training phase have been given in Table~\ref{NN_performance_ranges_8TeV}. The 
histogram in Figure~\ref{fig:13TeV_LHC_chi2_all} shows 
the difference between the CheckMATE and SCYNet results $\chi^2_{\rm
  SN}-\chi^2_{\rm CM}$. Again the comparison to the nearest neighbour interpolator shows that the neural network provides a much more powerful
  tool to predict the LHC $\chi^2$. The mean error
  for the neural network on points in the validation set is 1.45.
 
The performance of the 13~TeV network is also different for the
different $\chi^2$ ranges, see Figure~\ref{fig:13TeV_LHC_chi2_ranges}. However, the difference between the mean
error in the zero signal range and in the rare target ranges is less pronounced than for the
8~TeV network. This can be understood from comparing Figures~\ref{fig:chi2_distribution_8TeV}
and \ref{fig:chi2_distribution_13TeV}: the 13~TeV LHC analyses are
sensitive to a wider range of pMSSM-11 parameters, and thus fewer of
the sampled points result in no signal expectation.

\begin{figure*}[t]
\begin{subfigure}[b]{0.5\textwidth}
    \includegraphics[width=\textwidth]{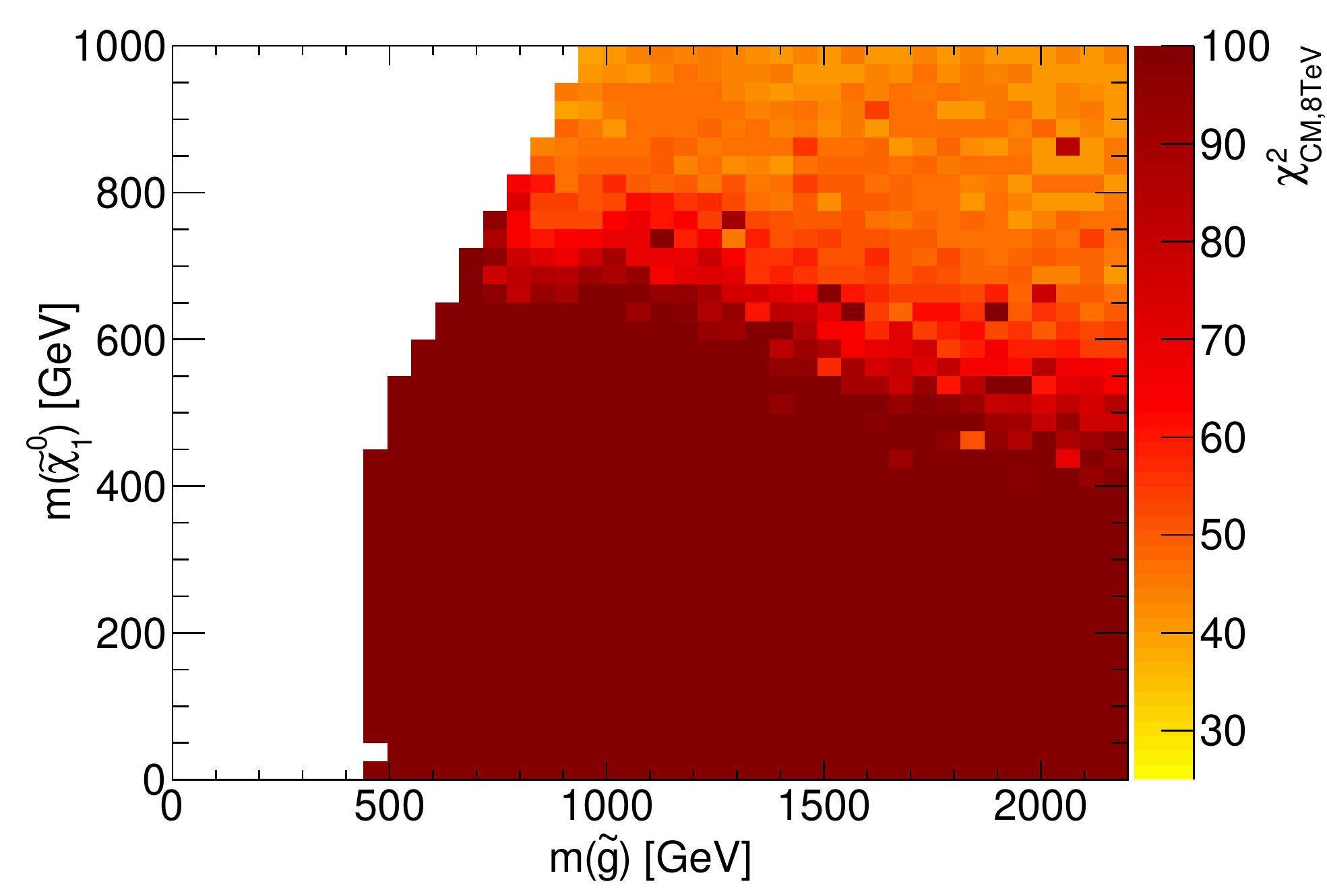}
    \caption{}
    \label{fig:profile_max_8TeV_mgo_mchi01_CM}
  \end{subfigure}
  \hfill
  \begin{subfigure}[b]{0.5\textwidth}
     \includegraphics[width=\textwidth]{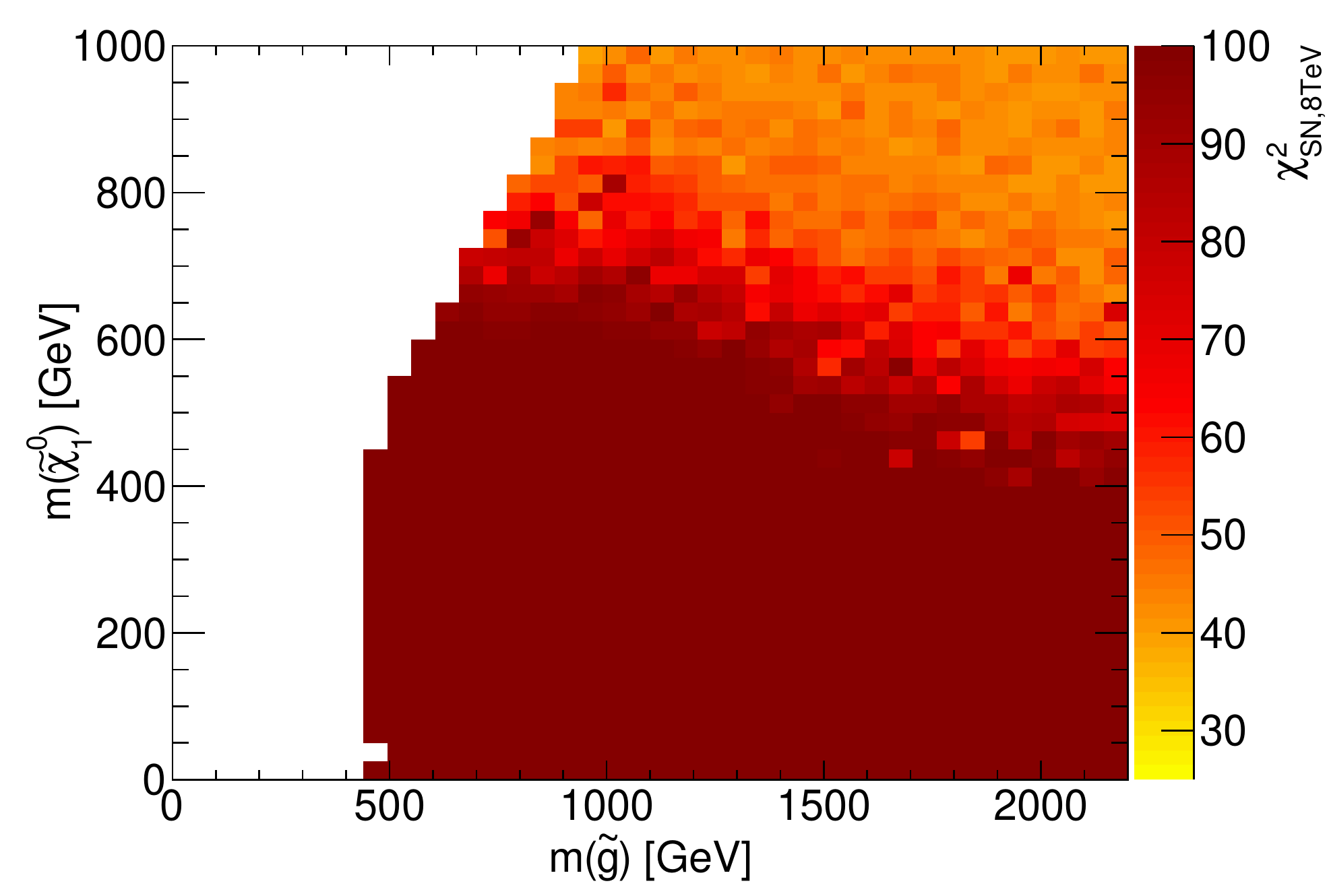}
    \caption{}
    \label{fig:profile_max_8TeV_mgo_mchi01_SN}
  \end{subfigure}
  \caption{\sl Maximum pMSSM-11 $\chi^2$ in the gluino-neutralino mass
    plane for the 8~TeV validation set for the CheckMATE
    results (a) and the predictions of SCYNet (b).}
\end{figure*}

\subsection{Testing the neural networks}

In this section we have used an additional, statistically independent validation set of 60000 
pMSSM-11 points that passed the preselection criteria to compare the
SCYNet prediction to the CheckMATE result. For illustration we have focused
on the projection of the 11-dimensional pMSSM parameter space onto the masses of the 
gluino $\tilde{g}$ and the neutralino $\chi^0_1$, which are
particularly relevant for the $\chi^2$ of the LHC searches. 
All plots in this section have been given for the 8\,TeV case, while similar results were obtained for the 13\,TeV case.

In Figure~\ref{fig:profile_min_8TeV_mgo_mchi01_CM} we have presented the
minimal $\chi^2_{\rm CM}$ obtained by CheckMATE for the
validation set of pMSSM-11 points in bins of the gluino and neutralino
masses. The minimum $\chi^2$ in each ($m(\tilde{g}),
m(\tilde{\chi}^0_1)$)-bin typically corresponds to scenarios,
where all other SUSY particles, and in particular squarks of the first two
generations, were heavy and essentially decoupled
from the LHC phenomenology. 

In Figure~\ref{fig:profile_min_8TeV_mgo_mchi01_SN} the
corresponding result obtained from the SCYNet neural network
regression has been given. We found 
that the neural network 
reproduces the main features of the $\chi^2$ distribution. We emphasize here that each bin
represents a single pMSSM-11 parameter point and consequently the results show that the network
successfully reproduces the LHC results across the whole plane. 

\begin{figure}[t] 
\begin{minipage}{\columnwidth}
\centering
\includegraphics[width=\textwidth]{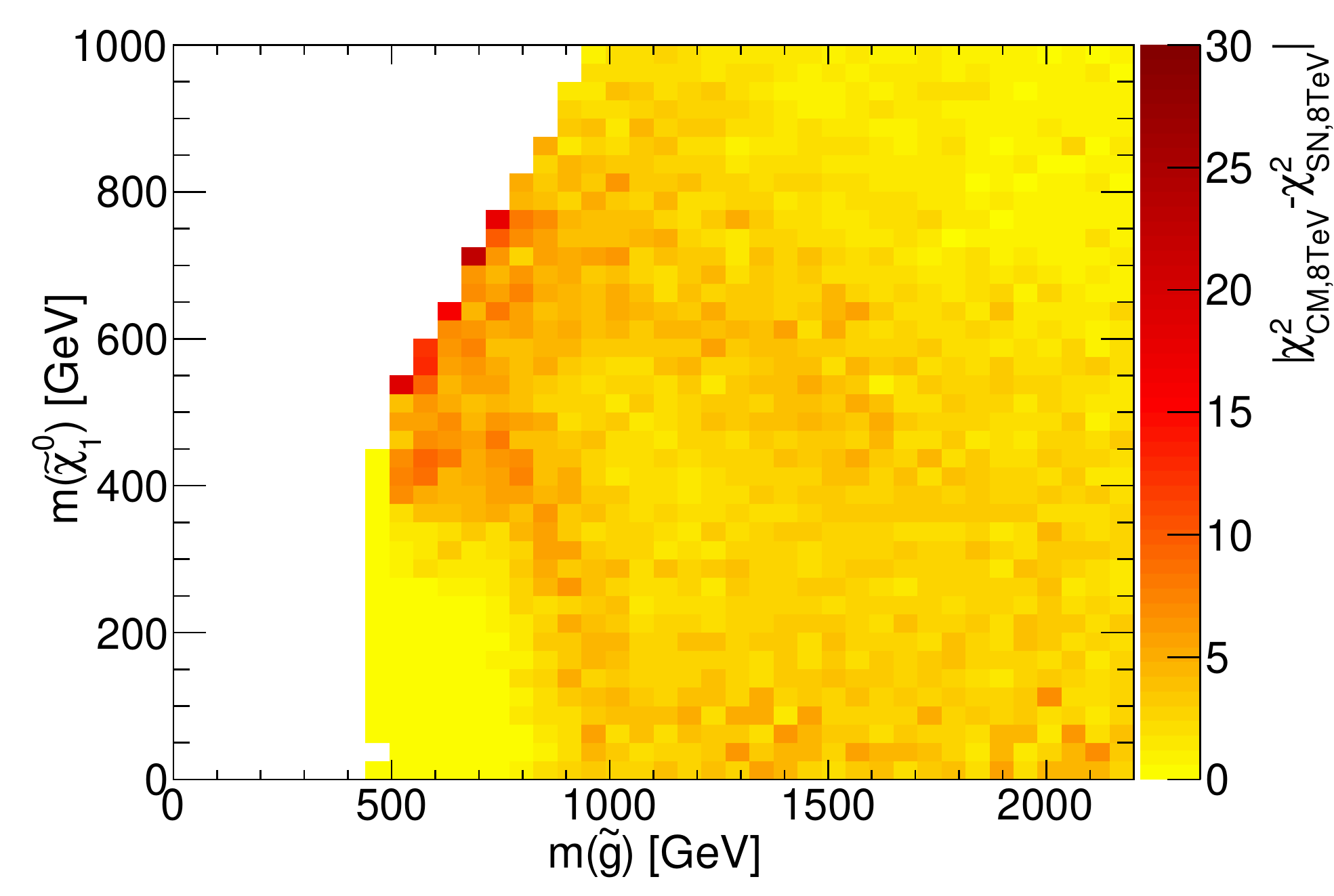}
\end{minipage}
\caption{\sl Difference between the CheckMATE and SCYNet $\chi^2$predictions in the gluino-neutralino mass plane for all 
8~TeV validation points. In each bin the mean difference has been calculated for all validation points.}
\label{fig:profile_8TeV_diff_CM_SN_mgo_mchi01}
\end{figure}

Figures~\ref{fig:profile_max_8TeV_mgo_mchi01_CM} and
\ref{fig:profile_max_8TeV_mgo_mchi01_SN} again show the $\chi^2_{\rm CM}$
and $\chi^2_{\rm SN}$ values as a function of the gluino and
neutralino masses, but now have we displayed the 
maximum $\chi^2$ in each bin. Comparing the two figures proves that
the neural network reproduces the main features of the $\chi^2$
distribution well.

\begin{figure*}[t]
  \begin{subfigure}[b]{0.5\textwidth} 
    \includegraphics[width=\textwidth]{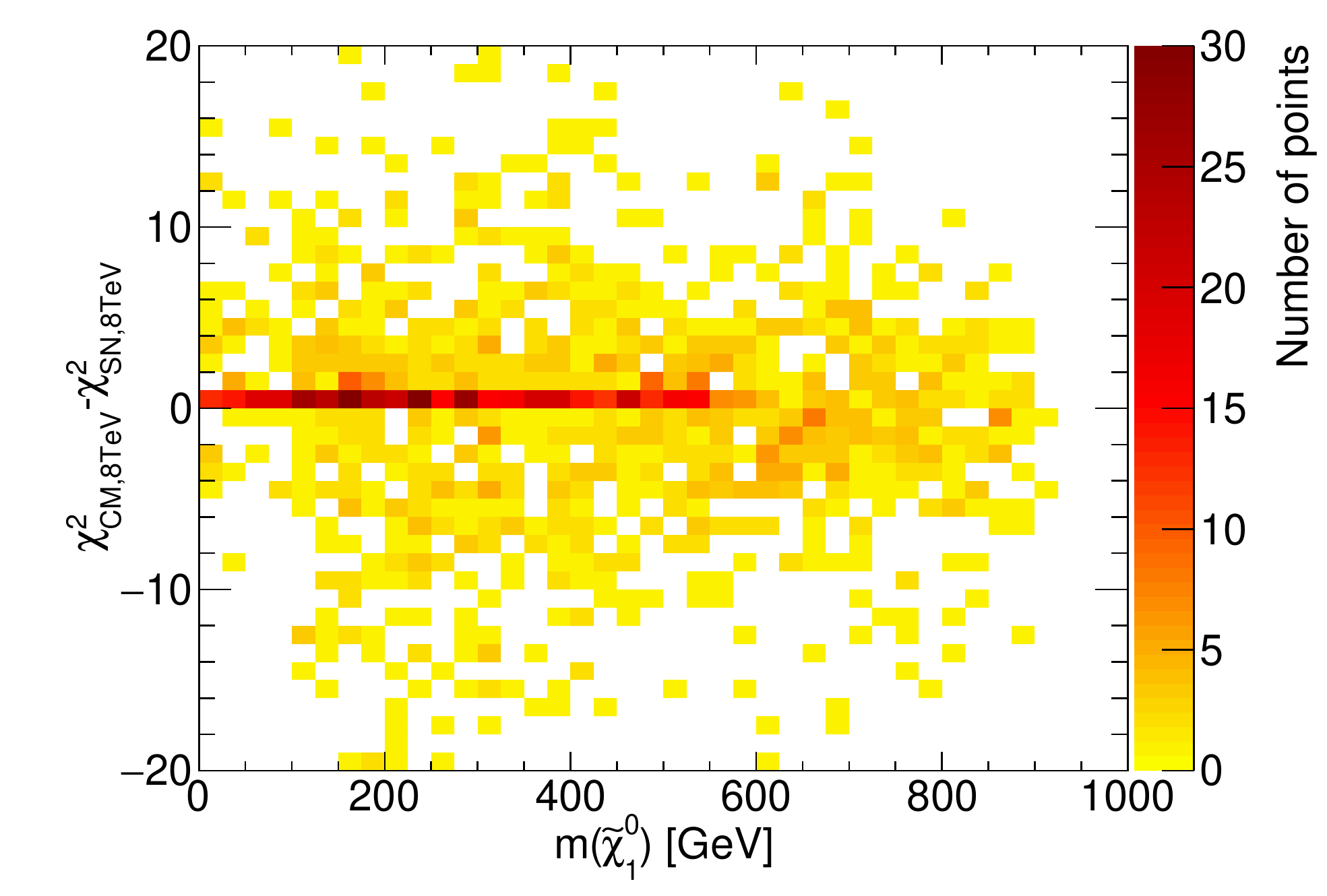}
    \caption{}
    \label{fig:profile_8TeV_diff_CM_SN_mgo_mchi01_Neutralinoprojection}
  \end{subfigure}
  \hfill
  \begin{subfigure}[b]{0.5\textwidth}
    \includegraphics[width=\textwidth]{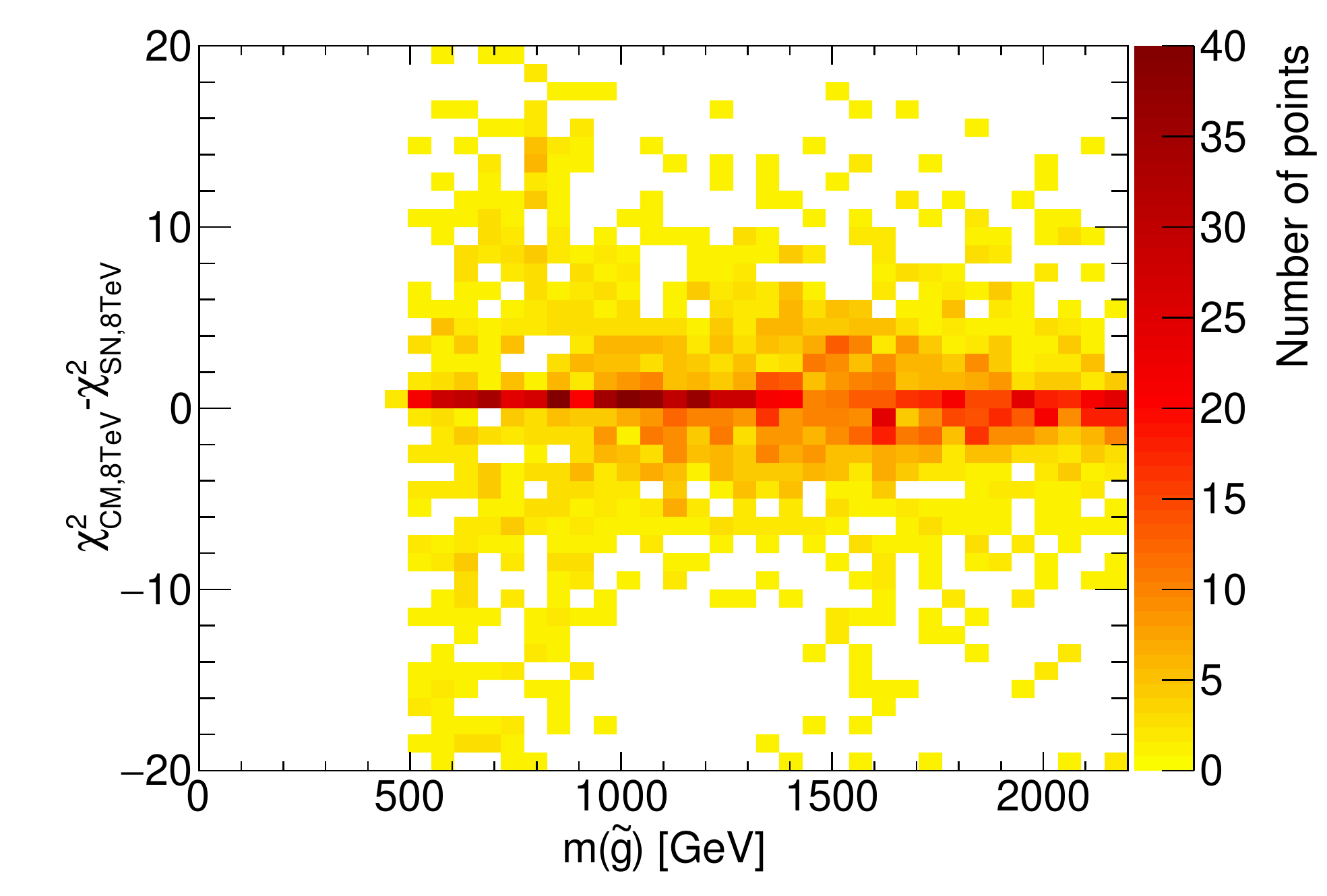}
    \caption{}
    \label{fig:profile_8TeV_diff_CM_SN_mgo_mchi01_Gluinoprojection}
  \end{subfigure}
  \caption{\sl Using all validation points from figure~\ref{fig:profile_8TeV_diff_CM_SN_mgo_mchi01}, we show 
  the difference between the SCYNet parametrization and the CheckMATE result against $m(\tilde{\chi}^0_1)$ for 
  gluino masses between 850 and 900 GeV (b) and against $m(\tilde{g})$ for neutralino masses between 400 and 450 GeV (b).}
\end{figure*}

The difference between the CheckMATE and SCYNet result,
$\chi^2_{\rm CM, 8\,TeV}- \chi^2_{\rm SN, 8\,TeV}$ has been presented in Figure~\ref{fig:profile_8TeV_diff_CM_SN_mgo_mchi01}. In 
this plot we have taken the mean difference between both results for all validation points that lie
in the respective histogram bin in order to demonstrate in which regions of parameter space the  
network performs best. We have found overall very good agreement between the CheckMATE and SCYNet
result. However,  sizeable differences were visible in the 
compressed regions where $m(\tilde{g})$ is close to
$m(\tilde{\chi}^0_1)$. This particular region has been probed by monojet 
searches \cite{ATLAS-1502-01518,ATLAS-1604-07773} which were sensitive
only for very degenerate spectra. Thus the $\chi^2$ contribution peaks suddenly as the
mass splitting between the SUSY states is reduced. Unfortunately, the rapid change in $\chi^2$ makes this region 
difficult for the neural network to learn and will be targeted specifically in future work by generating more training data
in this area.

As obvious from Figure~\ref{fig:profile_8TeV_diff_CM_SN_mgo_mchi01}, the parametrization of $\chi^2_{CM}$ through 
SCYNet worked very well on average. However, the pMSSM-11 points still have a 
rather broad distribution of $|\chi^2_{\rm CM}-\chi^2_{\rm SN}|$, especially near the crucial transition 
from the non-sensitive to the sensitive region. After all, this is
exactly the region where the \textit{rare target learning problem}
(RTLP) alluded to in section~\ref{sec:NN_training} occurs. 
In order to illustrate this, in Figure~\ref{fig:profile_8TeV_diff_CM_SN_mgo_mchi01_Neutralinoprojection} we have displayed the 
distribution of $\chi^2_{\rm CM}-\chi^2_{\rm SN}$ in bins of $m(\chi^0_1)$ for gluinos with a mass between 850 and 900 GeV. 
In Figure~\ref{fig:profile_8TeV_diff_CM_SN_mgo_mchi01_Gluinoprojection} the equivalent result for neutralino masses 
between 400 and 450 GeV has been given. The mass ranges were chosen such that we catch the transition regions from a low to 
a high $\chi^2_{\rm CM}$ in the minimum profile plot \ref{fig:profile_min_8TeV_mgo_mchi01_CM}. 
We can clearly say that in both cases there is a narrow peak around
$\chi^2_{\rm CM}-\chi^2_{\rm SN}=0$ and this is also true in 
the crucial transition regions. However we can also see that the RTLP causes few, but significant 
outliers, which will be subject to future improvements with targeted training.

We finally note that our results for the pMSSM-11  $\chi^2$ cannot be
compared in a straightforward way to exclusions published by
the LHC experimental collaborations. ATLAS and CMS have not presented
any specific analyses for the pMSSM-11, but typically interpret their searches in terms
of SUSY simplified models. The simplified models assume 100\,\%
branching ratios into a specific decay mode, which does not hold in
the pMSSM-11. Instead, in the pMSSM-11 there are in general a number
of competing decay chains that result in a large variation in the
final state 
produced in different events. As a result, the events do not predominantly fall into
 the signal region of one particular analysis, but are instead shared
 between many different analyses. Thus the constraints on the pMSSM-11
 are in general weaker than those on simplified SUSY models. 

\section{Neural network regression: reparametrized approach}\label{sec:reparameterized}

The direct approach of SCYNet discussed in section~\ref{sec:direct} allows for a successful representation of the 
pMSSM-11 $\chi^2$ from the LHC searches. However, despite the successful modeling, there is 
motivation to explore an alternative ansatz. Foremost, the model parameters used as input to the direct 
approach do not necessarily correlate with the $\chi^2_{\rm CM}$ behaviour.
For example, parameters such as $A$ and $\tan\beta$ are not directly linked to any single observable in the signal regions
considered here. This separation implies a complex function that has to be learned and modeled by the neural network itself. 
Another downside of the direct approach is that a neural network trained on the model parameters is inherently model dependent.  
For every model considered, new training data is required and a new neural network must be trained. 

In this section, a different set of phenomenologically motivated parameters was proposed as an input to a new network: 
The {\it reparametrized network}.  These input parameters are in principle observables, and more closely related to 
the $\chi^2_{\rm CM}$ values. Most importantly they are model independent. It was the aim of this ansatz to reach a 
performance which was at least comparable or better to the performance of the SCYNet direct approach. However, this 
comes at the cost of an increase in computation time, since for each evaluation, branching fractions and cross sections 
have to be calculated. The methods used for building and training the neural network were similar 
to the ones used in the direct approach but here we found that a deeper network of 9 layers performed better.

\subsection{Reparametrization procedure}
\label{ch:repar}

\begin{figure}[t]
\large
\begin{tikzpicture}[%
    >=triangle 60,              
    start chain=going below,    
    node distance=20mm and 40mm, 
    every join/.style={norm},   
    scale=0.8,
    every node/.style={transform shape},
    ]
\tikzset{
  base/.style={draw, on chain, on grid, align=center, minimum height=4ex},
  proc/.style={base, rectangle, text width=8em},
  proc2/.style={base, rectangle, text width=14em},
  proc3/.style={base, rectangle, text width=7em},
  proc4/.style={base, rectangle, text width=5em},
}
\node [proc3] (p0) {Model\\Parameters};
\node [proc3, below=of p0] (p1) {SLHA File};
\node [proc4, below left=20mm and 20mm of p1] (p2) {Particle 1};
\node [proc4, right=of p2] (p3) {Particle 2};
\node [below right=0mm and 15mm of p2] (p4) {\rotatebox[origin=c]{-90}{\includegraphics[width=.15\textwidth]{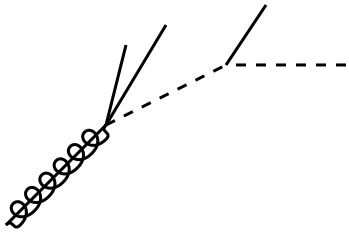}}};
\node [below left=0mm and 15mm of p3] (p5) {\reflectbox{\rotatebox[origin=c]{-90}{\includegraphics[width=.15\textwidth]{figures/feyn.png}}}};
\node[proc,right=35mm of p3,align=left] (lab1) {Consider final states of allowed \\ 2-to-2 production processes};
\node[proc,below=18mm of lab1,align=left] (lab2) {Consider all \\ decay chains};
\node [proc2, below=60mm of p1] (p6) {Observed particles and particle energies};
\node [proc2, below=of p6,align=center] (p7) {Observed particles and particle energies \\ across all decay chains};
\node [proc2, below=of p7,align=center] (p8) {Phenomenological parameters averaged across all decay chains and production cross sections};
\node[proc,below=25mm of lab2,align=left] (lab3) {Average decay chains weighted by their branching ratio};
\node[proc,below=25mm of lab3,align=left] (lab4) {Average production processes weighted by their cross sections};

\draw[->] (p0) -- node[right] {SPheno} (p1);
\draw[->] (p1.south) -- (p2.north);
\draw[->] (p1.south) -- (p3.north);
\draw[->] (p6.south) -- (p7.north);
\draw[->] (p7.south) -- (p8.north);
\end{tikzpicture}
\caption{\sl Flowchart of the reparametrization procedure. }
\label{fig:repar_flow} 
\end{figure}

A new model is excluded if in addition to the expected Standard Model background a statistically significant excess of events was
predicted for a particular signal region which however is not observed in data.  These observables generally 
correspond to a combination of final state particle multiplicity and their corresponding
kinematics. For example, if two distinct models produce the same observable final states 
at the LHC they would both be allowed/excluded independently of the mechanism producing the events. 
The reparametrized approach aims to calculate neural network inputs that relate more closely to these observables.  
As displayed in Figure~\ref{fig:repar_flow} all the allowed 2-to-2 production processes were considered.  
For each produced particle, the tree of all possible decays with a branching ratio larger than 1\% was
traversed. At the run time, each occurrence of a final state jet, b-jet, $e^{\pm}$, $\mu^{\pm}$, $\tau^{\pm}$, 
$\slashed{E}_{T}$ as well as intermediate state on-shell $W^{\pm}$,$Z$ and $t/\bar{t}$ were counted.  Note that 
the algorithm does not differentiate between the type of particle producing the missing energy since these are always
invisible to the detector. Therefore, the set of all 
invisible particle types (including for example neutrinos or a SUSY LSP) was considered as a single final state type. 
The charge conjugated partners of all final state particles and resonances (excluding jets, b-jets, $Z$ and $\slashed{E}_T$) 
were considered separately which adds to 9 separate final state categories and 5 different parameters for the resonances.
After all decay trees are constructed, the weighted mean and standard deviation of the number and the maximal 
kinematically allowed energy of observed final states particles and resonances were also calculated (see section~\ref{ch:phenopar} 
for more details). The weights were the individual occurrence probability of each respective decay tree.  Finally these 
quantities were averaged across all 2-to-2 production processes weighted by their individual cross sections.  

The aforementioned quantities complete the set of inputs in the reparametrized 
approach which adds up to a total of 56 parameters. The 
branching ratios were analytically calculated using SPheno-3.3.8~\cite{spheno} and read in with 
PYSLHA.3.1.1~\cite{Buckley:2013jua}. For strong processes NLL-Fast \cite{nllfast} was used for both 8 TeV and 13TeV
since the cross sections are quickly obtained from interpolations on a pre-calculated grid.  
For electroweak production processes Prospino2.1~\cite{prospino_1,prospino_2,prospino_3,prospino_4,prospino_5,prospino_6,prospino_7,Kramer:1997hh,Kramer:2004df,Alves:2002tj,Plehn:2002vy,Alves:2005kr} 
was used. In the standard configuration, the cross section evaluation of Prospino requires the most computing
time in our complete calculation. Consequently we only calculate the cross sections to leading order and we reduce
the number of iterations performed by the 
VEGAS~\cite{vegas} numerical integration routine, resulting in an increase on the statistical uncertainty from $<1\%$ to a 
few percent.
Obviously the induced error slightly damages the mean accuracy of the reparametrized neural net. Nevertheless, 
the final accuracy of the reparametrized approach was comparable to the direct approach, as demonstrated below. 

The total computation time of the reparametrization requires $3-11$ seconds and the Prospino run time was the dominant
factor despite the reduced integration accuracy. This was still a significant 
improvement over the ${\cal O}$(hours) needed without a
parametrization like SCYNet. However, compared to the order of millisecond
evaluation of the direct approach this makes the reparametrized approach less attractive for 
applications in global fits.

\subsection{Motivation and calculation of the phenomenological parameters}
\label{ch:phenopar}

Almost all analyses that search for new physics define the final state multiplicity (this can also be a
range) in the relevant signal regions. Consequently the mean number of final state
particles aims to help the neural net to find areas in parameter space where individual signal regions dominate. 
In addition, many analyses differentiate themselves by a kinematical selection of an on-shell $W$, $Z$ or $t$ 
in the decay chain, in order to isolate and exclude certain SM backgrounds. For this reason we have
also included the mean number of resonant SM states in the decay chains as another parameter.

When we examined the missing energy SUSY searches used in this study,
we found that the most prevalent cuts were those based on the energy of the final state particles.
This is also true of a missing energy cut which can be considered as the (vector) summation in 
the transverse plane of the energy of particles that go undetected.
Consequently we sought to introduce a parameter that is correlated to the energy of the 
particles that would be observed. However we had to keep in mind that the aim of our
study was to be able to calculate the LHC $\chi^2$ in as short a time as possible and any calculation
that consumes too much CPU, most importantly those which would require an integration over 
possible phase space configurations, disqualifies the reparametrization from being practical.

Since we already used the mean multiplicity of the various final state (and resonant) particles, the obvious 
choice would have been to calculate the mean energy for each of these states.
However, for a cascade decay that may have included a number of $1\to3$ and even $1\to4$ decays and such a calculation
requires a phase space integration, which unfortunately cannot be performed in the time frame available for each call to SCYNet.
As a replacement, we instead chose to use the maximum kinematically allowed energy a particle can have as a weighted
average over all production processes and possible decay chains.

\begin{figure*}[t]
\begin{subfigure}[b]{0.5\textwidth}
	\center
    \hspace{11pt}\includegraphics[trim={1.5cm 0 0cm 0},clip,width=0.9\textwidth]{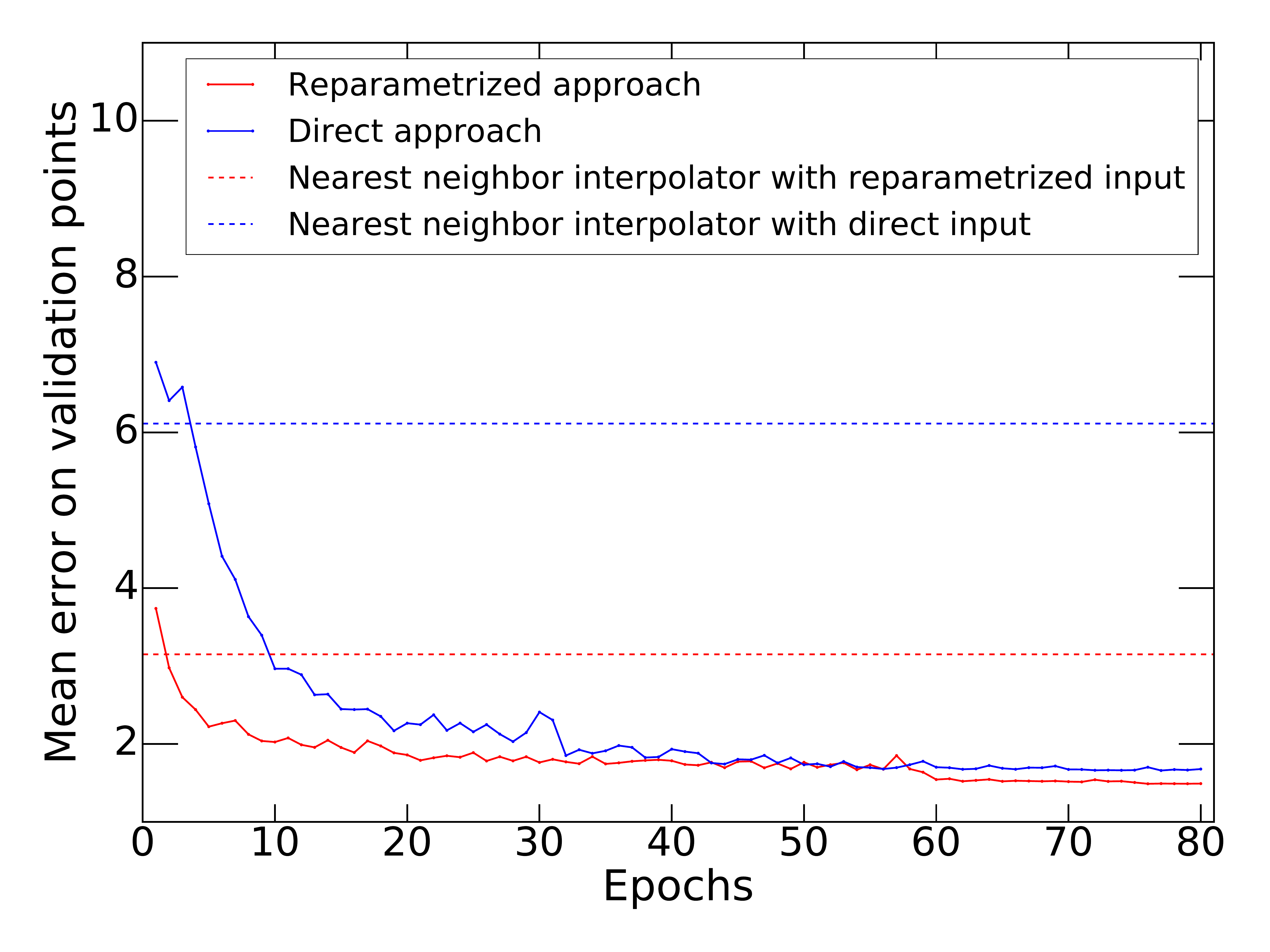}
    \caption{}
  \label{fig:repar_speed_a} 
  \end{subfigure}
  \hfill
  \begin{subfigure}[b]{0.5\textwidth}
	  \center
      \hspace{11pt}\includegraphics[trim={1.5cm 0 0cm 0},clip,width=0.9\textwidth]{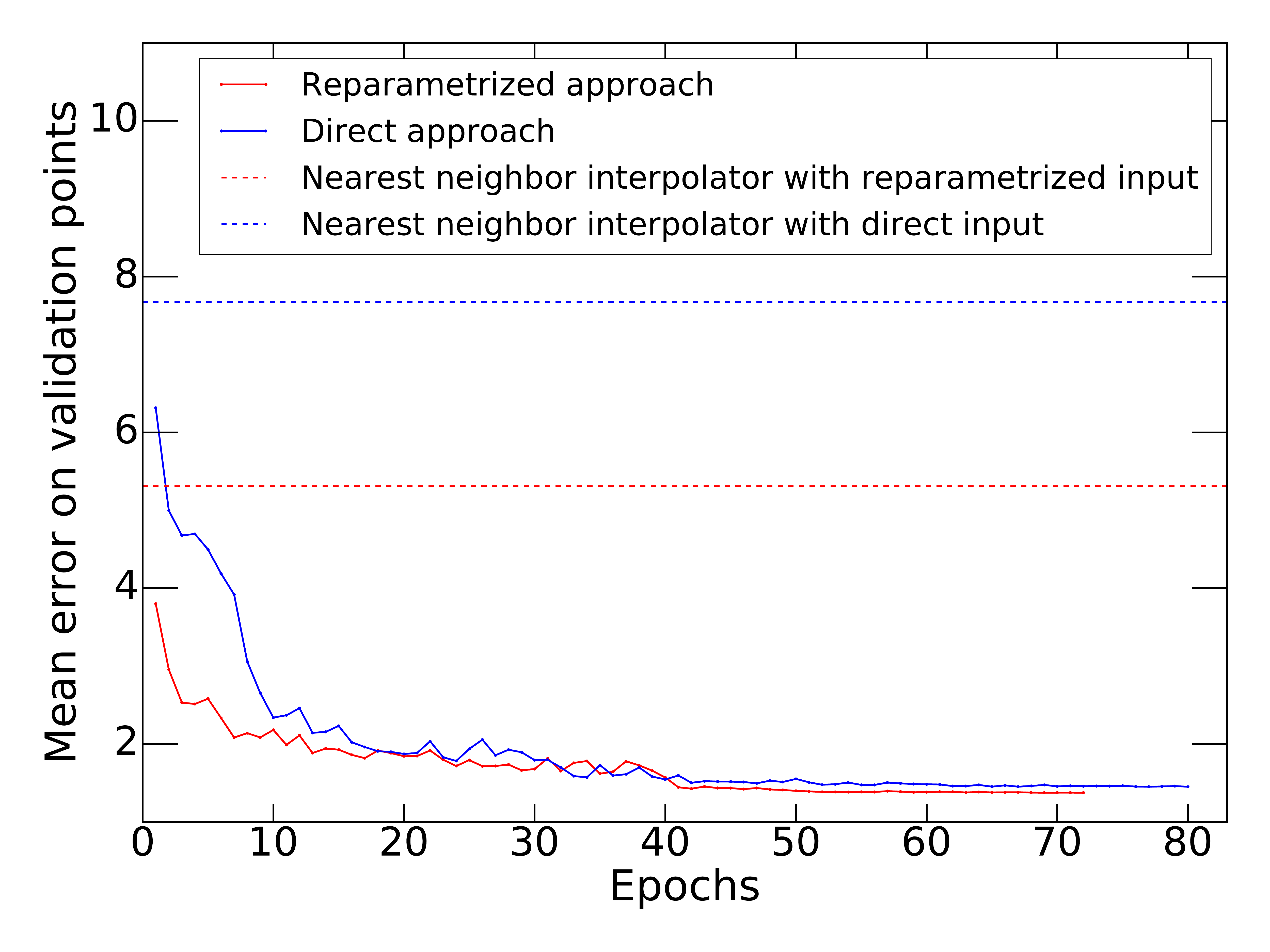}
    \caption{}
   \label{fig:repar_speed_b} 
  \end{subfigure}
  \caption{\sl Progression of the mean absolute error the direct and
    reparametrized neural networks achieve on the validation set after 
  each training iteration for 8 TeV (a) and 13 TeV (b). 
  }
\label{fig:repar_speed} 
\end{figure*}

In the center of mass frame, this maximum was calculated with the term,
\begin{equation}
E^{\text{max}}_{i}=\sqrt{\frac{(m_k^2-(\sum_j m_j + m_i)^2)(m_k^2-(\sum_j m_j-m_i)^2)}{(2 m_k)^2}+m_i^2}\,.
\end{equation}
$E^{\text{max}}_{i}$ gives the maximum energy of daughter particle $i$ with mass $m_i$ in the decay  $k \rightarrow i + \{j\}$ in 
the center-of-mass of the mother $k$. $m_k$ is the mass of the mother
particle while $m_j$ represents
the mass of the other daughter particle(s) $j$.
This term is fast to calculate and the same for any number of daughter particles.

In the next step, the energy of $i$ must be boosted from the center of mass frame of the mother $k$
into the lab frame:
\begin{align}
E^{\text{max, Lab}}_{i} = \gamma_{k}\; E^{\text{max}}_{i}\,.
\end{align}
Here, the boost $\gamma_{k}$ depends on the energy of the mother $k$ as measured 
in the lab-frame which we have estimated at an earlier stage in the decay 
chain, $\gamma_k = E^{\text{Lab}}_k/m_k \approx E^{\text{max, Lab}}_{k}/m_k$. For the
initial mother (heaviest) particle in the decay chain, we assumed that 
this was produced at rest and consequently, $E^{\text{max, Lab}}_{\text{initial}}$ was fixed to $m_{\text{initial}}$.

\bigskip

During averaging across all decay trees, not only the mean of the parameters but also the standard 
deviation was calculated and used as an input. The motivation for such a parameter can easily be understood
if for example we consider two new physics scenarios, one which always produces a single lepton
in the final state, while the other produces four leptons but only in 25\% of decay chains. Both
of these scenarios have a mean number of leptons equal to one but if we have a signal region
that requires 4-leptons, clearly only the second scenario will satisfy such an analysis.

The set of inputs in the reparametrized approach added 56 parameters but it cannot be 
trivially assumed that the function mapping the 
presented parameters to $\chi^2_{\rm SN}$ is injective, nor that they are uncorrelated.  In a way, 
the parameters can be seen as a replacement for a first layer of the neural net preparing the 
inputs to facilitate the modelling process. Their viability should be evaluated based on the 
results they produce.  In theory the reparametrization could be performed for any model, but 
in practice the cross sections and branching ratios must be obtainable in a timely manner 
which is why we have restricted ourselves to supersymmetric models where such tools were already
available.

\subsection{Architecture and training of the reparametrized neural network}

\begin{figure*}[t]
\begin{subfigure}[b]{0.5\textwidth}
    \hspace{11pt}\includegraphics[trim={1.5cm 0 0cm 0},clip,width=0.9\textwidth]{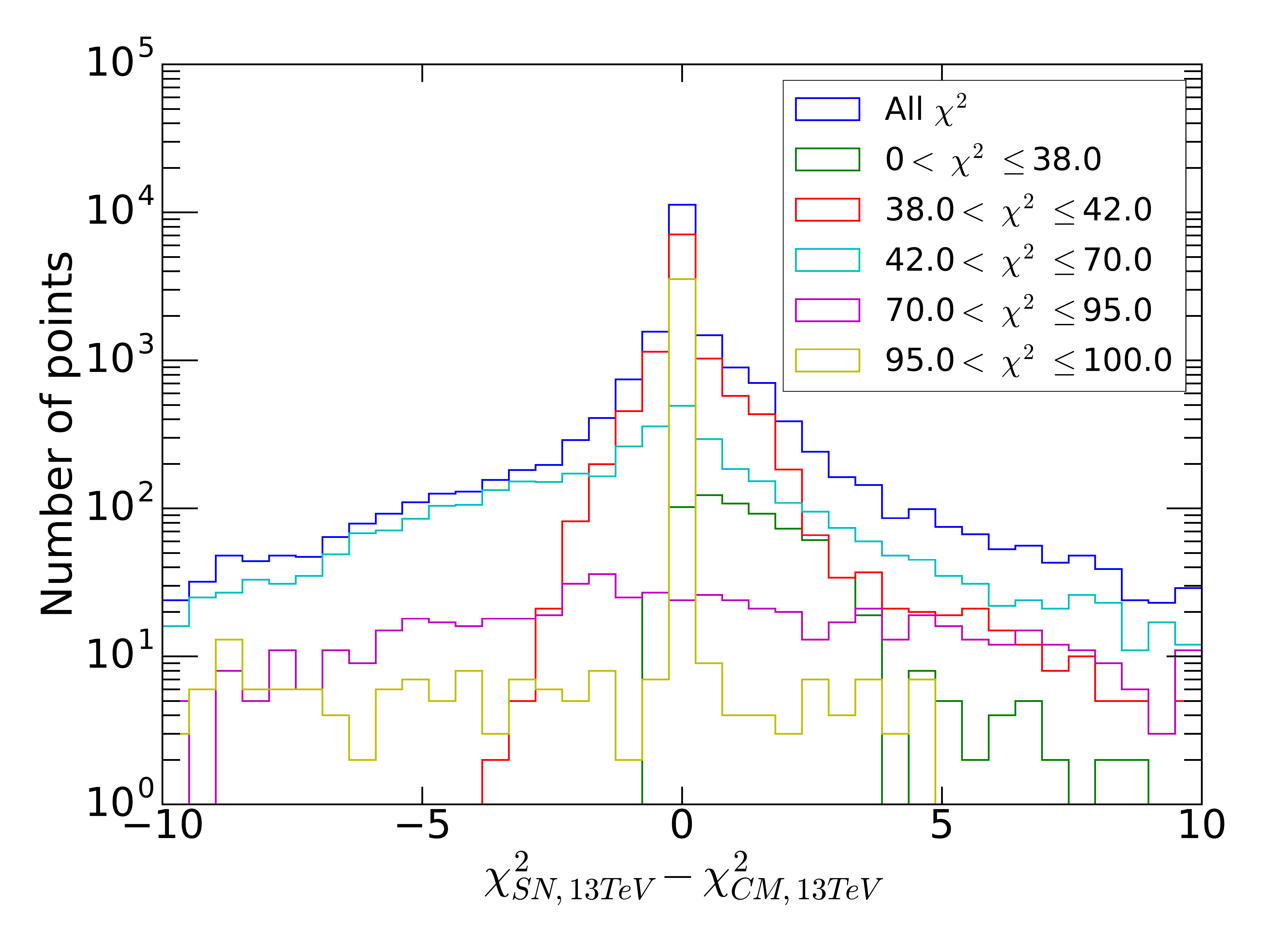}
    \caption{}
    \label{fig:repar_pull_a}
  \end{subfigure}
  \hfill
  \begin{subfigure}[b]{0.5\textwidth}
    \hspace{11pt}\includegraphics[trim={1.5cm 0 0cm 0},clip,width=0.9\textwidth]{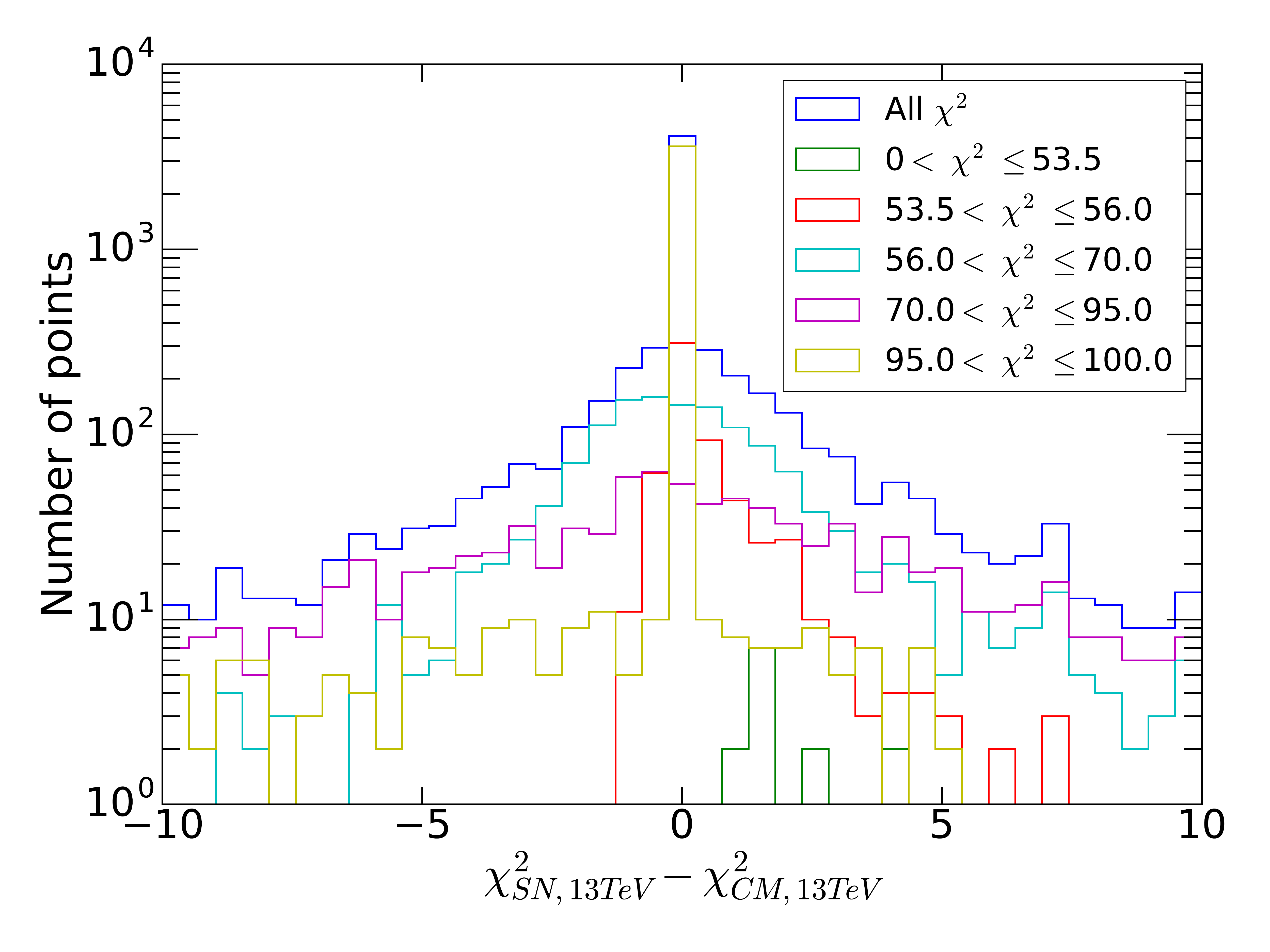}
    \caption{}
    \label{fig:repar_pull_b}
  \end{subfigure}
  \caption{\sl Performance of the neural network trained on the reparametrized pMSSM-11 points at 8~TeV (a) 
  and 13~TeV (b). }
  \label{fig:repar_pull}
\end{figure*}

\begin{table}[t!]
\centering
\begin{tabular*}{\columnwidth}{@{\extracolsep{\fill}}lll@{}}
\hline
Energy in TeV & Direct approach & Reparameterized approach\\
\hline
8  & 1.7~ (3.1 \%) & 1.5~ (2.7 \%)\\
13 & 1.5~ (2.0 \%) & 1.3~ (1.8 \%)\\
\hline
\end{tabular*}
\caption{\sl Mean errors on the LHC $\chi^2$ prediction for both approaches in SCYNet.}
\label{tab:NN_performance}
\end{table}

For the reparametrized approach, we found that a deeper architecture was preferred with a first layer built from 500 hidden nodes 
followed by 8 layers of 200 hidden nodes.
The added depth, creates additional challenges and required us to alter the hyperparameter setup that we used in the direct approach.
Here, rectified linear units (relu)~\cite{relu} were used as activation functions due to them being less prone to the vanishing gradient problem.
Due to the larger architecture, overfitting was expected to play a larger role and to counteract this, the regularization
term was increased by a factor of 100 to be $\lambda=0.001$, see Eq.~(\ref{eq:cost_func}). In addition the batch size was 
reduced to 120 (32 for 13TeV) to improve convergence.\footnote{The added random element supports the minimizer in overcoming flat areas and local minima.} 
As in the direct approach, before starting the training, the inputs were again forced to undergo a 
Z-score normalization. Since the relu activation function is not restricted, the modification described in \ref{sec:modified_z_score_normalization}
was not necessary. 

In contrast to the model parameters, the reparametrized inputs were heavily correlated and consequently the inputs were decorrelated by projecting 
the data into the eigenvectors of the covariance matrix. These mappings were calculated based on the training 
dataset and then stored to be able to apply the same mappings when evaluating the network. 
For the training itself, the nAdam~\cite{adam,nadam1,nadam2} optimizer was used. Compared to the Adam
optimizer used in the previous sections, the nAdam optimizer adds Nesterov 
momentum.\footnote{Instead of applying the momentum with the gradient, the momentum 
is applied first and the gradient is calculated after the update for the updated weights.}
The learning rate was initialized a higher value of 0.01 for 8\,TeV and 0.002 for 13\,TeV. Similar to the batch size, 
the larger learning rate improved the convergence behaviour in the early stages of the training and was later removed since, as in the direct approach, 
the learning rate was reduced if the mean error has stopped improving.

\begin{figure*}[t]
\begin{subfigure}[b]{0.5\textwidth}
    \includegraphics[width=\textwidth]{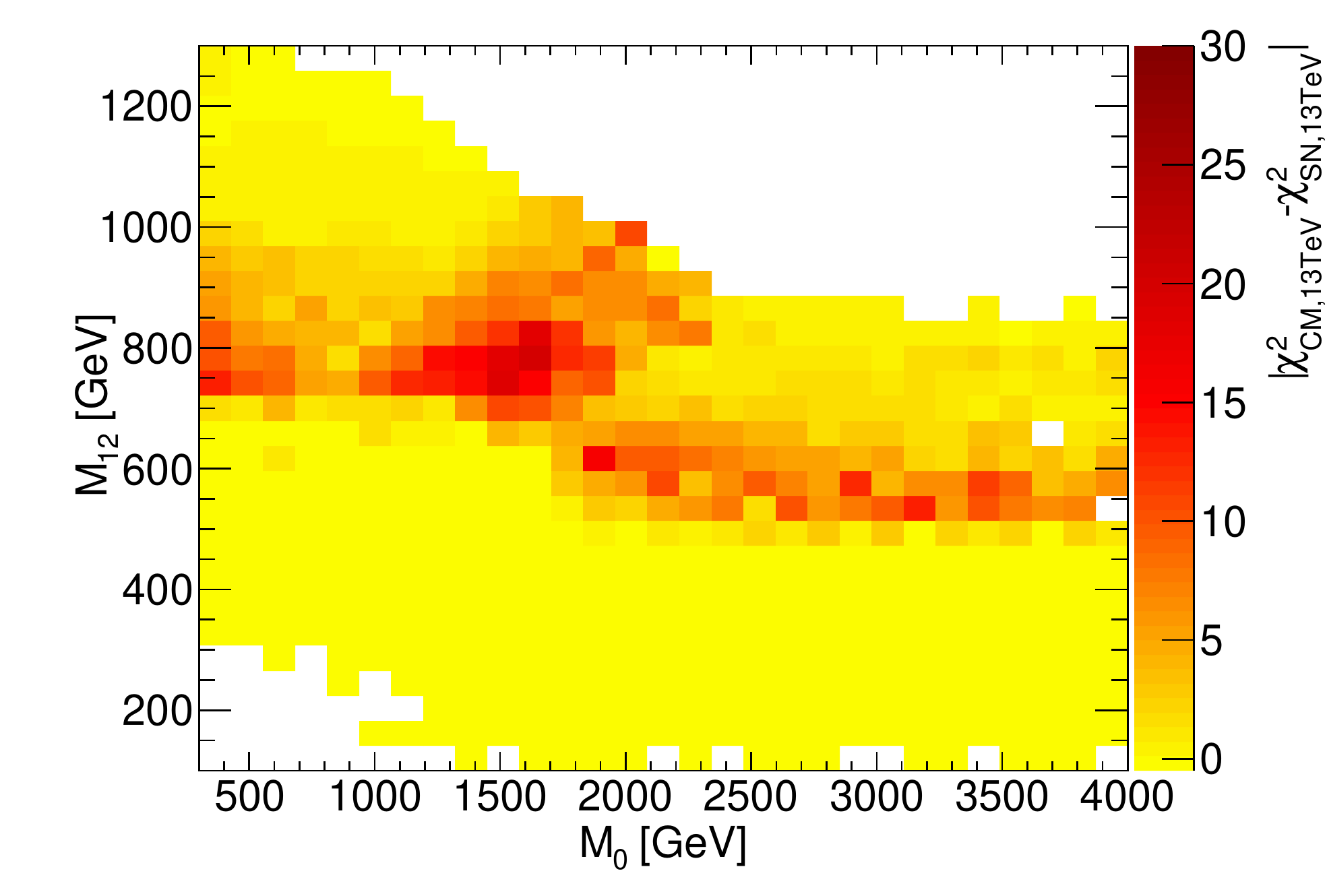}
    \caption{}
    \label{fig:repar_heat_a} 
  \end{subfigure}
  \hfill
  \begin{subfigure}[b]{0.5\textwidth}
    \includegraphics[width=\textwidth]{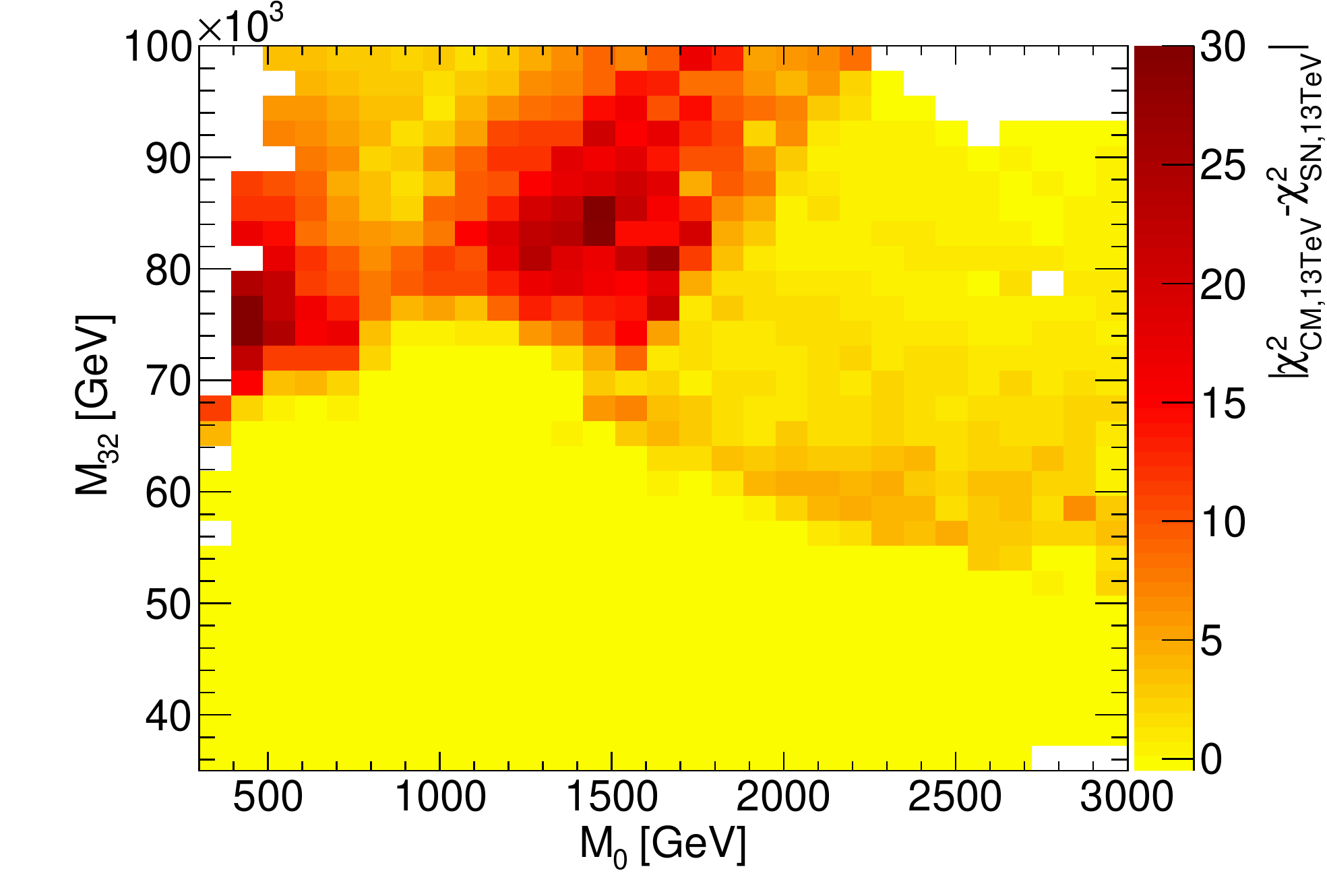}
    \caption{}
    \label{fig:repar_heat_b} 
  \end{subfigure}
  \caption{\sl Performance of the neural network trained on the reparametrized pMSSM-11 points on the cMSSM (a) and AMSB (b). 
  In each bin the mean difference was calculated for all validation points.}
 \label{fig:repar_heat} 
\end{figure*}

\subsection{Results of the reparametrized approach to LHC neural nets}

During the calculation of the phenomenological parameters, cross sections and branching ratios were calculated. 
These quantities are related to the expected signal events in each signal region in a much more direct way than the model parameters. 
Consequently the target quantity of the parametrization, the $\chi^2$ derived from the signal counts in each non-overlapping signal region, 
should be more directly dependent on the input parameters of the
reparametrized network. 
This hypothesis was supported by Figure~\ref{fig:repar_speed} which showed that a nearest neighbor interpolator improves 
(in both the 8~TeV and 13~TeV case) when moving from the direct to the reparametrized approach. We interpreted this result 
as showing that the function mapping the reparameterized inputs to the outputs was flatter.

This improvement directly relates to the performance of neural nets which
is given in Table~\ref{tab:NN_performance} as the total mean error of all networks 
while the distribution of the mean error has been given in Figure~\ref{fig:repar_pull}.
In total, the reparameterized approach on average displayed a lower error. 
Consequently, as we have shown in Table~\ref{NN_performance_ranges_8TeV}, 
the reparametrized approach outperformed the direct approach in several $\chi^2$ ranges. 
This is offset by the significantly longer computation time required for each call of the reparametrized net in SCYNet. 
The dominant component of this computation is the time required by Prospino to calculate the electroweak cross sections.
Additional research is now being performed to see if this calculation
can be done by a separate neural network. That may allow to bring down
the computation time as to be competitive 
with the direct approach.

The most useful upside of the reparametrized approach was that the used input parameters were 
chosen in a way which should make them (in principle) model independent. This could 
allow a network which was trained with one model to be used as prediction tool for a 
wide variety of models. In Figure~\ref{fig:repar_heat} we have displayed the performance of the 13~TeV 
reparametrized network trained with the pMSSM-11 samples when fed with points from the cMSSM and 
the AMSB. We emphasize here that due to Renormalization Group Equation (RGE) running, both the 
cMSSM and AMSB models are {\it not}
subsets of the pMSSM-11 since in general all scalar masses become non-degenerate. In large 
regions of the parameter space, the network has predicted the model $\chi^2$ directly calculated 
from CheckMATE correctly. However, along the transition region from clearly excluded (in the 
bottom of the frame) to clearly not excluded (in the top of the frame) a few regions of parameter
space display discrepancies. 

\begin{table*}[t]
\centering
\begin{tabular*}{0.8\textwidth}{@{\extracolsep{\fill}}llll@{}}
\hline
Model & \parbox[t]{5cm}{Only pMSSM-11 samples} & \parbox[t]{5cm}{Additional pMSSM-19 samples} \\
\hline
pMSSM-11  & 1.32 & 1.50 \\
CMSSM & 2.43 & 1.92 \\
AMSB & 2.44 & 1.63\\
\hline
\end{tabular*}
\caption{\sl Performance of the 13~TeV neural network in the reparametrized approach applied to models different from the pMSSM-11 used for training.}
\label{repar:model_independence_table}
\end{table*}

\begin{figure*}[t]
\begin{subfigure}[b]{0.5\textwidth}
    \includegraphics[width=\textwidth]{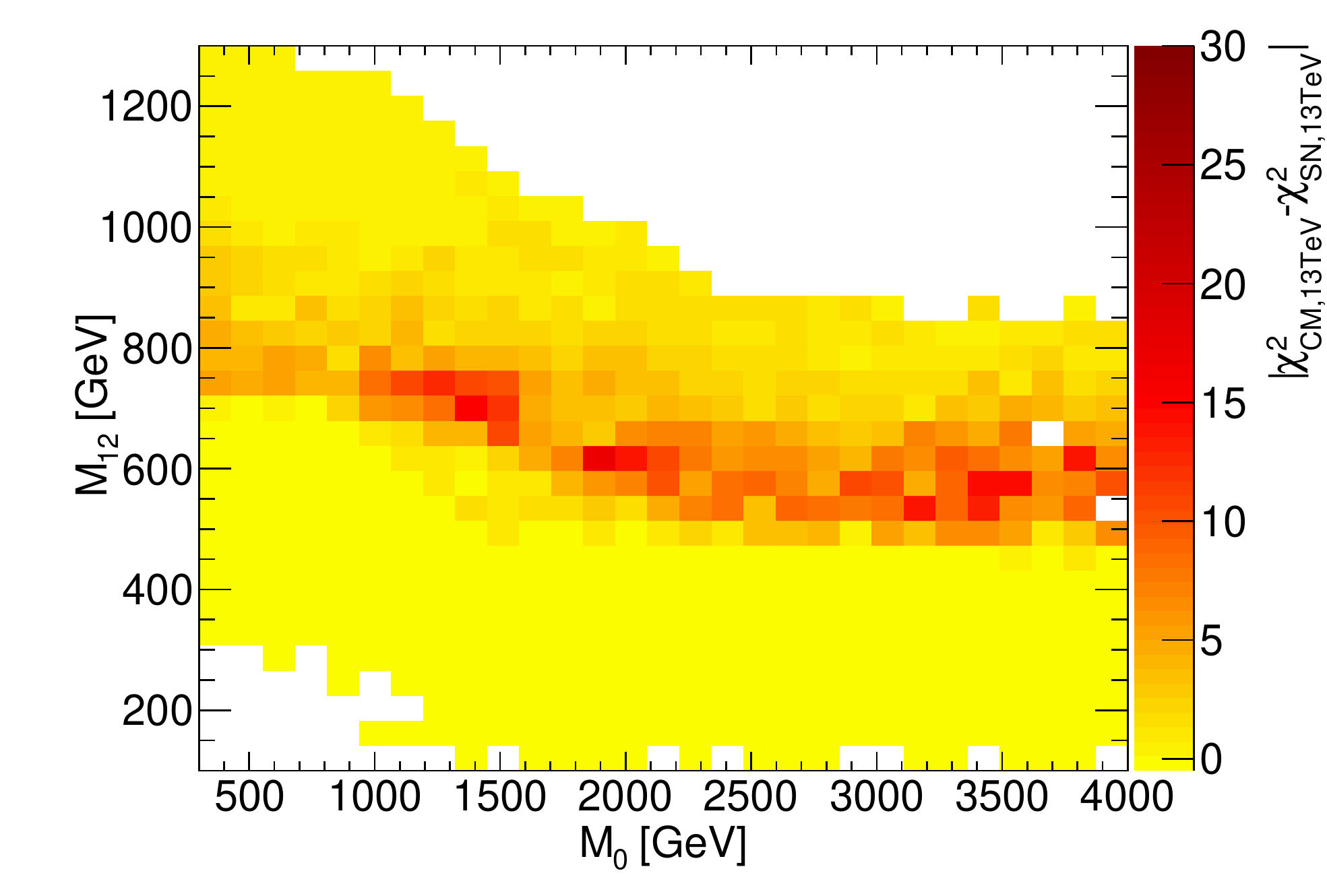}
    \caption{}
  \label{fig:repar_heat_extra_a}  
  \end{subfigure}
  \hfill
  \begin{subfigure}[b]{0.5\textwidth}
    \includegraphics[width=\textwidth]{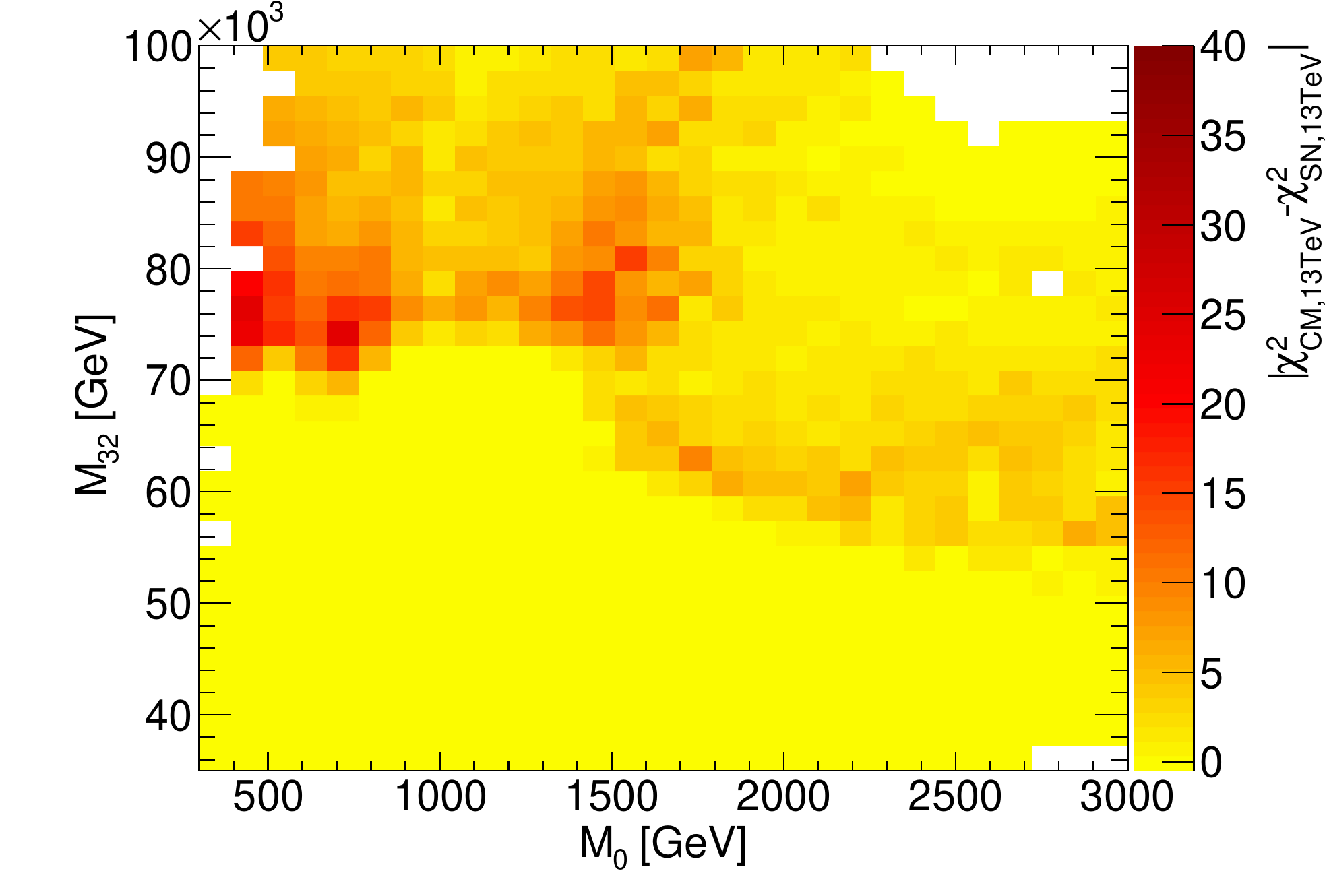}
    \caption{}
  \label{fig:repar_heat_extra_b}  
  \end{subfigure}
  \caption{\sl Performance of the neural network trained on the reparametrized pMSSM-11 points with added samples 
  containing non-degenerate stop sbottom pairs on the cMSSM (a) and AMSB (b). 
  In each bin the mean difference was calculated for all validation points.}
  \label{fig:repar_heat_extra}  
\end{figure*}

In particular a visible anomaly can be seen in the top left of Figure~\ref{fig:repar_heat} 
for the AMSB and to a lesser degree the cMSSM. These wrongly predicted points arise because 
in the pMSSM-11 model, all the stop and the sbottom states (both left- and right-handed) were assumed to be 
mass degenerate. However, as already alluded to, RGE effects mean that in general this was not
the case in the AMSB and cMSSM. In fact, the model points in the 
anomaly region exhibited a light gluino decaying exclusively into a stop-top pair,
\begin{align}
 \tilde{g} \to \tilde{t}\, t\,,
\end{align}
since the sbottoms are heavier than the gluino here. The stops then further decay, producing
either another top quark and a neutralino or a b-quark and a chargino depending on the exact details of the mass spectra via,
\begin{align}
 \tilde{t} \to  t\, \tilde{\chi}^0 \quad\quad\text{or} \quad \quad \tilde{t} \to b\, \tilde{\chi}^{\pm}\,.
\end{align}
Further decays of the top/chargino always produce a $W$-boson, leading to two $W$-bosons per decay chain
and four in the complete event.
In the considered pMSSM-11 model however,
one always additionally observes the gluino decay via an on-shell sbottom and these often do not 
decay via $W$-Bosons. The average number (and standard deviation) of
resonant $W$-Bosons is part of the neural network 
inputs in the reparametrized approach. Since all decays were averaged during the reparametrization 
procedure, the pMSSM-11 samples will never contain such a large number
of intermediate $W$ bosons. Thus, when we apply the neural network to the AMSB or cMSSM, we no longer interpolate
between known parameter points but instead extrapolate to regions of parameter space that
the network has never sampled as they were not represented in the training data. 

A simple solution to counteract the lack of training data that represents parameter points that
only contain light gluinos and stops was to generate additional points 
in the pMSSM-19 \cite{Djouadi:1998di}. 
For these points, the particles charged under SU(3) were forced to obey
the following mass relation,
\begin{equation}
m_{\tilde{t}}<m_{\tilde{g}}<m_{\tilde{u},\tilde{d},\tilde{s},\tilde{c},\tilde{b}} 
\end{equation}
in order to generate training data in the region the reparameterized network fails.

The results of a network trained with the additional pMSSM-19 samples have been displayed in 
Table~\ref{repar:model_independence_table}. The network performs slightly worse on the 
pMSSM-11 set which was to be expected because the net had to focus on regions of the 
parameter space which were not present in the pMSSM-11. On the other hand, the mean 
error of the network applied to the AMSB and cMSSM was significantly reduced.  The effects 
of this can be observed in Figure~\ref{fig:repar_heat_extra} which has significantly reduced 
errors compared to Figure~\ref{fig:repar_heat}. 

One may notice that anomalous areas still exist with larger
errors. This was due to the fact that 
we did not sample the additional pMSSM-19 points with a high enough density for the neural
network to learn the parameter space properly. The points with the worst reconstruction again
contain a spectrum with a lighter stop and can be expected to be
improved if the pMSSM-19 points were sampled correctly from the beginning.

\section{Conclusions}\label{sec:conclusion}

We have developed a neural network regression approach, called SCYNet, for calculating  
the $\chi^2$ of a given SUSY model from a large set of 8\,TeV and 13\,TeV LHC  searches. Previously, such a $\chi^2$ calculation 
would require computational intensive and time consuming Monte Carlo simulations. The SCYNet neural network 
regression, on the other hand, allows for a fast $\chi^2$ evaluation and is thus well suited for 
global fits of SUSY models. 

We have explored two different approaches: in the first, so-called direct, method we simply used the pMSSM-11 parameters as input to the neural network. 
Within this method the $\chi^2$-evaluation for an individual pMSSM-11 parameter point only takes few milliseconds and can thus be 
used for global pMSSM-11 fits without 
time penalty. However, the neural network based on the pMSSM-11 parameters cannot be used 
for any other model. In the second approach, we reparametrized the input to the neural network. Specifically, we used the 
SUSY masses, cross sections and branching ratios of the pMSSM-11 
to estimate signature-based objects, such as particle energies and multiplicities, which provide a more model-independent input 
for the neural network regression. Although calculating 
the particle energies and multiplicities for any given parameter point requires ${\cal O}({\rm seconds})$ of computational time, the reparametrized network trained on a 
particular model can in principle be applied to BSM scenarios the network has not encountered before. 

The mean error of both neural network approaches lies in the range 
of $\Delta\chi^2=1.3 - 1.7$, corresponding to a relative precision between 2\% and 3\%. This is
already very close to the accuracy required for a global fit, where the profile likelihood 
equivalence between 1\,sigma and $\Delta\chi^2=1$ implies $\Delta\chi^2 < 1$ as an appropriate goal in precision. For such a complex application with ${\cal O}(50)$ non-overlapping 
signal regions, the SCYNet implementation represents the most advanced neural network regression for the pMSSM-11 to date.

Applying the reparametrized pMSSM-11 neural network to the cMSSM and AMSB models, we found a few regions of 
parameter space where the network would fail to predict the correct $\chi^2$. As the cMSSM and AMSB models are not subsets of the pMSSM-11, there are 
cMSSM and AMSB parameter configurations where the network has to extrapolate to regions that
were not represented in the training data. We have, however, demonstrated 
that such problems can be addressed systematically by additional specific training of the network.

Our results motivate the continuation and further improvement of the neural network regression approach 
to calculating the LHC $\chi^2$ for BSM theories. 
A more accurate approximation to the true LHC likelihood can, for example, be obtained by modeling the probability density 
functions used for the event generation of the training sample to increase the sampling density in 
specific regions of parameter space. Furthermore, the treatment of systematic correlations between signal regions should 
be studied in more detail. The direct approach should be extended to other models, such as the pMSSM-19, so that the neural 
networks can be used in the corresponding global fits, and the reparameterized neural network regression should be studied further 
in the context of additional models.

With these improvements in mind, it is possible to decrease the mean error 
of the SCYNet neural network approach significantly below $\Delta\chi^2\approx-2\ln{\cal L}<1$, thus 
making it a powerful tool for a large variety of global BSM fits.

\begin{acknowledgements}
  The work has been supported by the German Research Foundation (DFG)
  through the Forschergruppe \textsl{New Physics at the Large Hadron
    Collider} (FOR 2239), by the BMBF-FSP~101 and in part by the
  Helmholtz Alliance ``Physics at the Terascale''. We also would like
  to thank Jong Soo Kim, Sebastian Liem and Roberto Ruiz de Austri for 
  discussions in the early stages of this project.
\end{acknowledgements}

\appendix

\section{$\chi^2$ calculation for one single SR}\label{chi2_SR}\label{appendix:chi2}

     In this section we explain how to calculate the profile likelihood ratio ($PLR$) for one SR which can be interpreted as the '$\chi^2$' value used in text.\\

      For a given signal region, the following results are available
\newcommand{\sig}{N_S}
\newcommand{\bkg}{N_{\mathrm{SM}}}
\newcommand{\obs}{N_{E}}
\newcommand{\dSsys}{\sigma_{\sig}^{\mathrm{sys}}}
\newcommand{\dSstat}{\sigma_{\sig}^{\mathrm{stat}}}
\newcommand{\dBsys}{\sigma_{\bkg}^{\mathrm{sys}}}
\newcommand{\dBstat}{\sigma_{\bkg}^{\mathrm{stat}}}
\newcommand{\dS}{\sigma_{\sig}}
\newcommand{\dB}{\sigma_{\bkg}}
\newcommand{\nuS}{\nu_S}
\newcommand{\nuB}{\nu_{\mathrm{SM}}}
            \begin{itemize}
            \item Number of predicted signal events $\sig$,
            \item statistical and total systematic error on signal events $\dSstat$, $\dSsys$,
            \item number of predicted SM events $\bkg$, summed over all background sources, 
            \item statistical and total systematic error on SM events $\dBstat$, $\dBsys$ and
            \item number of experimentally observed events $N_{E}$.
            \end{itemize}
            At first one constructs the likelihood as follows
            \begin{equation}
            \mathcal{L}(\obs|\mu,\nuS,\nuB):=\frac{e^{-\lambda}}{\obs!}\lambda^{\obs} \cdot\frac{1}{\sqrt{2\pi}} e^{-\frac{\nuS^2}{2}} \cdot \frac{1}{\sqrt{2\pi}} e^{-\frac{\nuB^2}{2}}. \label{eq:appstat:likeli}
            \end{equation}
            The first term originates from the Poisson distribution which describes the compatibility of observing $\obs$ events if $\lambda$ are expected. Here, $\lambda$ --- which itself is a function of the parameters $\mu, \nuS, \nuB$ explained below --- is given as follows:
            \begin{equation}
            \lambda(\nu_S,\nu_{SM},\mu) := \mu \sig\ e^{\frac{\dS}{\sig}\nuS}+ \bkg\ e^{\frac{\dB}{\bkg}\nuB}. 
            \label{eq:appstat:lambda}
            \end{equation}
The total uncertainties on signal and background 
\begin{eqnarray}
\dS &:= \sqrt{\left(\dSstat\right)^2+\left(\dSsys\right)^2}, \\
\dB &:= \sqrt{\left(\dBstat\right)^2+\left(\dBsys\right)^2}
\end{eqnarray}
are reparameterized in terms of dimensionless, Gaussianly distributed nuisance parameters $\nuS, \nuB$ in Eq.~(\ref{eq:appstat:likeli}). Eq.~(\ref{eq:appstat:lambda}) then describes lognormal\footnote{A lognormally distributed variable is asymptotically Gaussian in the limit $\Delta X \ll X$ but forbids unphysical negative values for $\Delta X / X = \mathcal{O}(1)$.} distributions of $\sig, \bkg$ with respective widths $\dS, \dB$.

            In a stochastic picture, we now wish to compare the null hypothesis $H_0$ and the alternative hypothesis $H_1$:
            \begin{eqnarray}
            H_0:\ \ \parbox{18em}{The predicted number of signal events, i.e.\ $\mu = 1$, coincides well with the experimental observation}, \\
\nonumber \\
            H_1:\ \ \parbox{18em}{Any number of signal events, i.e.\ any $\mu$, coincides well with the experimental observation}.
            \end{eqnarray}
            To decide optimally between these two hypotheses, the Ney\-man-Pearson lemma \cite{10.2307/91247} proposes the profile likelihood ratio
            \begin{eqnarray}
            PLR := \frac{\mathcal{L}_C}{\mathcal{L}_G},
            \end{eqnarray}
            with the respective profile likelihoods defined as
            \begin{eqnarray}
            \mathcal{L}_C &:=& \max_{\nu_S, \nu_{SM}\in \mathbb{R}}\ \mathcal{L}(\mu=1,\nu_{SM},\nu_S), \\
            \mathcal{L}_G &:=& \max_{\mu,\nu_S, \nu_{SM}\in \mathbb{R}}\ \mathcal{L}(\mu,\nu_{SM},\nu_S). 
            \end{eqnarray}
            Here, the constrained likelihood $\mathcal{L}_C$ maximizes $\mathcal{L}$ with respect to the nuisance parameters for the null hypothesis with fixed $\mu=1$. In contrast, the global likelihood $\mathcal{L}_G$ also varies $\mu$ to find the global maximum of $\mathcal{L}$. Note that the range of allowed $\mu$ is not restricted here and both negative as well as values beyond unity are allowed.\footnote{Note that this is different to e.g.\ limit setting procedures where $\mu$ is restricted to values $\leq 1$.}

            According to Wilk's theorem \cite{wilks1938}, the variable
            \begin{equation}
            q_{\mu} := -2 \log PLR
            \label{negative_log_profile_likelihood}
            \end{equation}
            is asymptotically $\chi^2$ distributed in the limit of large $\obs$. We can validate this statement for our setup in the simplified case of negligible uncertainties $\dS$, $\dB$. Then, following the prescription above, one finds $q_\mu = (\sig + \bkg - \obs)^2/(\sig + \bkg)$ which is indeed the $\chi^2$ distribution for one degree of freedom with observation $\obs$ and expectation $\sig + \bkg$.

We therefore refer the value of $q_\mu$ whenever we use the expression '$\chi^2$' within this work.

\section{Modified Z-score normalization}\label{sec:modified_z_score_normalization}
The $\chi^2$s which are our targets are called
\begin{equation}
y_1,\cdots y_N \in \mathbb{R}^+
\end{equation} 
in this appendix.
We apply the Z-score normalization to them
\begin{equation}
\hat{y}_i=\frac{y_i-\mu}{\sigma}.
\end{equation}
Furthermore we define
\begin{equation}
\hat{y}_{max}:=\max \hat{y}_i,~~~y_{max}:=\max y_i
\end{equation}
and finally we normalize again
\begin{equation}
\hat{\hat{y}}_i=\frac{\hat{y}_i}{\hat{y}_{max}} \in (-1,1).
\end{equation}
All targets are therefore in (-1,1).\\
Usually one chooses $\mu=\frac{1}{N}\sum y_i$ and $\sigma=\frac{1}{N-1}\sum (y_i-\mu)^2$ but this would cause 
problems with the back transformed outputs of the network. We want the back-transformed outputs of the 
network to be between the maximum  $z_1=100$ and the minimum $z_0=$ minimum possible LHC $\chi^2$.\\
With the choice
\begin{equation}
\mu=\frac{z_0+y_{max}}{2}
\end{equation}
$z_0$ corresponds to an output value of $\hat{\hat{y}}= -1$. \\
$z_1$ corresponds to an output value of $\hat{\hat{y}}=+1$. \\
Therefore we choose,
\begin{equation}
   \sigma=\frac{1}{N-1}\sum\limits_{i} (y_i-(\frac{1}{N}\sum\limits_j y_j))^2
\end{equation}

\bibliographystyle{spphys}
\bibliography{neural_net_paper}

\end{document}